\documentclass[twocolumn]{aastex701}

\shorttitle{Extended Blue Light of NGC\,1275}
\shortauthors{Levitskiy et al.}

\begin{document}


\title{Tidal Disruption of Super Star Clusters as the origin of Bluish Light at the inner region of the \\ Perseus Cluster Central Galaxy}

\author{Arsen Levitskiy}\email{arsen15@connect.hku.hk}
\affiliation{Department of Physics, The University of Hong Kong, Pokfulam Road, Hong Kong}
\affiliation{Center for Astrophysics and Supercomputing, Swinburne University, John Street, Hawthorn VIC 3122, Australia}

\author{Youichi Ohyama}\email{ohyama@asiaa.sinica.edu.tw}
\affiliation{Academia Sinica, Institute of Astronomy and Astrophysics, 11F of Astronomy-Mathematics Building, No.1, Section\,4, Roosevelt Rd., Taipei 106319, Taiwan, R.O.C.}

\author{Jeremy Lim}\email{jjlim@hku.hk}
\affiliation{Department of Physics, The University of Hong Kong, Pokfulam Road, Hong Kong}



\begin{abstract}

Relatively blue light extends beyond a spiral disk from a radius of $\sim$5\,kpc\ out to $\sim$14\,kpc from the center of NGC\,1275.Analyses of its spectrum and broadband colors reveal a population of young stars having sub-solar metallicities superposed on a dominant population of old stars having super-solar metallicities.  The young stars have a characteristic age of $\sim$160\,Myr and may span ages of a few hundred Myr, similar to that of stars comprising the central spiral disk, and a total mass about one-third that of this disk for a  combined stellar mass (at birth) of $\sim$$4 \times 10^9 \rm \, M_\sun$.  A multitude of arc-like features embedded in the extended blue light have brightnesses comparable to the somewhat older ($\sim$500\,Myr) super star clusters (SSCs) projected against the central spiral disk.  The SSCs have a relatively shallow mass function, suggesting that the tidal disruption of an initially larger population that we estimate could have had an initial total mass (far) exceeding $\sim$$1 \times 10^9 \rm \, M_\sun$ gave rise to the extended blue light -- the arc-like features corresponding to stellar streams tracing disrupted star clusters -- and perhaps also the central spiral disk.  We speculate that, beginning about 500\,Myr ago, an enhanced episode of AGN activity in NGC\,1275, leaving still visible X-ray bubbles, induced vigorous cooling of the intracluster medium to fuel the formation of numerous star clusters: many were tidally disrupted to leave bluish light at the inner region of this galaxy, with the survivors being the SSCs projected against the central spiral disk.




\end{abstract}


\section{Introduction} \label{sec:Intro}

In the local universe, there must surely be few galaxies more visually enigmatic and more heavily scrutinized than NGC\,1275, the giant elliptical galaxy at the center of the Perseus Cluster.  This galaxy first rose to prominence for its association with a bright radio source as discovered by \citet{Baade1954}, listed in the Third Cambridge Catalogue of Radio Sources \citep{Edge1959} as 3C\,84.  As the brightest radio source in the Perseus constellation, 3C\,84 also is known as Perseus\,A.  NGC\,1275 gained further prominence when \citet{Minkowski1957} presented spectroscopic measurements showing optical line-emitting gas at two (radial) velocities differing by $3000 \, \rm km \, s^{-1}$ towards NGC\,1275, from which he concluded NGC\,1275 to be a colliding system comprising ``a tightly wound spiral of early-type and a strongly distorted late-type spiral''.  By contrast, \citet{Burbidge1963} and \citet{Burbidge1965} subsequently attributed the more redshifted line-emitting gas to an outburst in the nuclear region of NGC\,1275.  Over a decade later, the nature of NGC\,1275 remained a puzzle: in a paper titled `The NGC\,1275 Enigma', \cite{VandenBergh1977} concluded that ``NGC\,1275 is not composed of a normal late-type galaxy superimposed on (or interacting with) a giant elliptical.  No alternative interpretation, that is consistent with all presently-available observations, suggests itself."

\begin{figure*}[hbt!]
\centering
\includegraphics[width=\textwidth]{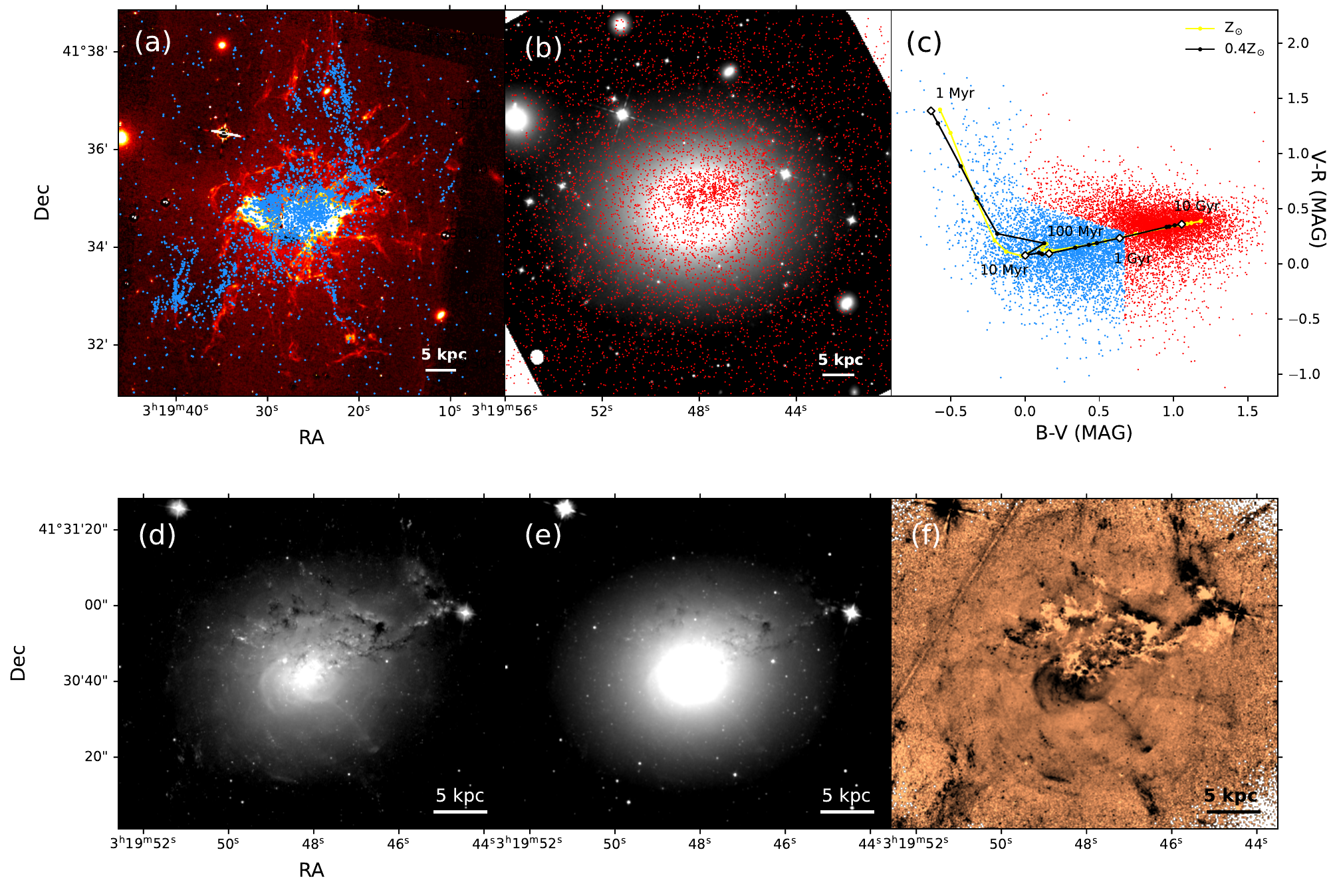}
\caption{Different features of NGC 1275 as observed with the Hubble Space Telescope.  (a) Blue dots indicate individual blue star clusters (BSCs) cataloged and identified by \citet{Lim2020} as progenitor globular clusters, superposed on an image in H$\alpha$+[NII] tracing an optical emission-line nebula.  (b) Red dots indicate individual old globular clusters (GCs), except for the concentration north to north-west of center associated with the High-Velocity System (HVS), also cataloged by  \citet{Lim2020}, superposed on a $H$-band image.  (c) Color-color diagram of the BSCs and GCs, overlaid on which are theoretical evolutionary tracks for star clusters having a Kroupa initial mass function and metallicities of $0.4 \, Z_{\odot}$ (black) or $1.0 \, Z_{\odot}$ (yellow) \citep[taken from][]{Lim2020}. (d) $B$-band image. (e) $H$-band image. (f) Unsharp masked $B-H$ image, highlighting bluish (darker shade) arc-like structures associated with the extended blue light and bluish spiral arms associated with the central spiral disk.  The HVS spans north to north-west of the centre of NGC\,1275, appearing redder (lighter shade) when it imposes especially strong dust extinction on NGC\,1275 and bluish (darker shade) along star-forming regions in its spiral arms.   }
\label{fig:NGC1275}
\end{figure*}

Since then, numerous studies have contributed to unraveling the mystery of NGC\,1275.  The optical line-emitting gas at a lower (radial) velocity originates from a filamentary nebula associated with NGC\,1275.  The full splendor of this nebula, showcasing its complexity, was not revealed until the photograph by \citet{Lynds1970}; a more modern ground-based image is presented by \citet{Conselice2001}, and an image with the Hubble Space Telescope (HST) by \citet{Fabian2008}.  In Figure\;\ref{fig:NGC1275}$a$, we show an image of this nebula (not fully covering its northern portion) taken with the HST.  The line-emitting gas at a higher (radial) velocity originates from a distorted late-type spiral, as originally proposed by \citet{Minkowski1957}.  Dubbed the High Velocity System (HVS), this galaxy is associated with silhouette dust, as shown in the $B$- (F435W) and $H$-band (F160W) images of Figure\;\ref{fig:NGC1275}$d$--$e$, and must therefore be located in front (projected towards the north-west) of NGC\,1275.  From an analysis of the observed absorption of intracluster X-ray-emitting gas by the HVS, \citet{Sanders2007} find that this galaxy must be about 110\,kpc from NGC\,1275.  At this distance from and at its velocity with respect to NGC\,1275, it seems unlikely that the distorted morphology of the HVS is caused by tidal interactions with NGC\,1275.  Indeed, \citet{Yu2015} showed that the HVS is experiencing strong ram-pressure stripping on what its likely its first passage through (on falling into) the Perseus cluster, thus explaining its distorted morphology.

The nebula associated with NGC\,1275 is multiphase, comprising not just atomic (which may be mixed with partially ionized) gas as revealed in optical emission lines, but also fully ionized gas as detected in X-rays \citep{Fabian2011}, along with molecular gas as detected in the 1-0~S(1) ro-vibrational transition of molecular hydrogen \citep[H$_2$;][]{Lim2012} as well as through rotational transitions of carbon monoxide \citep[CO;][]{Lazareff1989, Mirabel1989, Reuter1993, Braine1995, Inoue1996, Bridges1998, Salome2008, Salome2011, Lim2008}.  Such a nebula is not unique to NGC\,1275, but also seen in the central giant elliptical galaxies of other clusters \citep[e.g.,][]{Heckman1989,Crawford1999,Hamer2016} having relatively low core entropies as inferred for their intracluster medium (ICM) \citep[][]{Cavagnolo2008,Levitskiy2024} -- providing circumstantial evidence that these nebulae originate from cooling of the ICM.  The physical conditions that lead to thermal instabilities in the ICM, however, remain a subject of intensive debate \citep[e.g.,][]{Voit2015, Voit2017, McNamara2016, Hogan2017}.

NGC\,1275 exhibits a multitude of relatively young, compact, and massive star clusters, resembling globular clusters except for their youth.  The first such star clusters were discovered by \citet{Holtzman1992} towards the inner region of NGC\,1275, constituting the most luminous star clusters in this galaxy.  Recently, \citet{Lim2022} presented a detailed study of these star clusters that they referred to as super star clusters (SSCs), so as to distinguish them (for reasons that we shall return to) from the more numerous star clusters discovered farther out by \citet{Carlson1998}.  Subsequent studies of the latter star clusters by \citet{Conselice2001}, \citet{Canning2010}, and \citet{Canning2014} showed that those farther out are closely associated with, but mostly not spatially coincident with, the emission-line nebula in NGC\,1275.  More recently, \citet{Lim2020} cataloged all the compact stars clusters in NGC\,1275: they showed that the relatively young star clusters, which they referred to as blue star clusters (BSCs), have a complex spatial distribution as shown in Figure\;\ref{fig:NGC1275}$a$, by comparison with the more homogeneous spatial distribution of (old) globular clusters as shown in Figure\;\ref{fig:NGC1275}$b$. Furthermore, \citet{Lim2020} showed that the BSCs have formed at an approximately constant rate over the past, at least, $\sim$1\,Gyr, as shown in Figure\;\ref{fig:NGC1275}$c$.  Beyond this age, they cannot be distinguished from the globular clusters present in even greater numbers around NGC\,1275.  Remarkably, the BSCs have a mass function similar to the globular clusters \citep[see Fig.\,3 of][]{Lim2020}, suggesting that the former comprise progenitor globular clusters.

Despite being first remarked upon by \citet{Minkowski1957}, relatively little attention had been paid until recently to the remarkable spiral-like structure at the inner region of NGC\,1275, as can be seen in the $B$-band image of Figure\;\ref{fig:NGC1275}$d$.  This structure spans a radius of $\sim$5\,kpc, the same radius over which the especially luminous star clusters discovered by \citet{Holtzman1992} are distributed.  \citet{Yeung2022} showed that this spiral-like feature comprises spiral arms superposed on a roughly circular disk in rotation about the center of NGC\,1275:\,\,stars that make up this spiral disk have ages (if coeval) of between 100\,Myr and 200\,Myr, and a total mass of $\sim$3 $\times$ $10^9 \rm \, M_\sun$.  By contrast, as shown by \citet{Lim2022}, the SSCs associated with the central spiral disk -- those discovered by \citet{Holtzman1992} -- span a narrow range of ages of $500 \pm 100 \rm \, Myr$, older than the stars that make up the spiral disk.  \citet{Yeung2022} propose that the spiral disk formed from gas that originated from cooling of the ICM.  \citet{Lim2022} argue the same for the SSCs, which formed first followed by the formation of the spiral disk a few hundred Myr later. 

Beyond the central spiral disk, a multitude of arc-like features can be seen as described by \citet{Conselice2001,Penny2012} and shown in the unsharp masked $B-H$ image of Figure\;\ref{fig:NGC1275}$f$.  Whereas \citet{Penny2012} attribute these features to remnants of a cannibalized galaxy, \citet{Lim2022} suggest that these features may be stellar trails of tidally-disrupted SSCs.  Such tidal disruptions would provide a natural explanation for why the SSCs have a flatter mass function than the BSCs \citep{Lim2022}, as star clusters having lower masses are more easily disrupted than those having higher masses, as are those located closer to the center of NGC\,1275 where the tidal field is stronger.  
In the following, we address this last mystery that is apparent in optical images of NGC\,1275: the extended blue light, composed in part of a multitude of arc-like features, extending beyond the central spiral disk out to a radius of $\sim$14\,kpc, as is most clearly seen in Figure\,\ref{fig:slits}$a$. 
This image was made by subtracting the light of the old stellar population from the young stellar population in NGC\,1275, following the procedure described in Section\,\ref{sec:extendedcolors}. 

\citet{Romanishin1987} was the first to point out a steep radial color gradient for NGC\,1275 that implies much bluer colors at its inner regions, opposite to that normally displayed by (massive) elliptical galaxies owing to a metallicity gradient (higher metallicity and hence redder color inwards).  \citet{Romanishin1987} attribute this extended blue light to ongoing star formation at a rate of a few tens of $\rm M_\sun \, yr^{-1}$ at the inner region of NGC\,1275.
Similarly, \citet{McNamara1996} attribute the extended blue light at the inner region of NGC\,1275 to stars having ages over the range $\sim$10\,Myr--1\,Gyr that formed through cooling of the ICM (see also \citet{McNamara1992} and \citet{Rafferty2008} for discussion on color gradient of NGC 1275 as part of larger sample of cooling-flow clusters).  In this work, we show that the blue light extending beyond the central spiral disk is produced by stars having a similar age, or spanning a similar range of ages, as stars that make up the spiral disk.  We propose a new scenario that unites the formation of the extended blue light, central spiral disk, and SSCs associated with the spiral disk -- along with the mechanism that possibly triggered the formation of all these structures, uniting them with cavities in the ICM (X-ray bubbles) seen around NGC\,1275 created by relativistic jets from the active galactic nucleus (AGN) in this galaxy.
Throughout this manuscript, we adopt $H_{0} = 70 \rm \, km \, s^{-1} \, Mpc^{-1}$, for which the distance to NGC\,1275 is $\sim$ 74 Mpc.  At this distance, $1\arcsec = 360 \rm \, pc$.

\section{Data} \label{sec:DataProc}

\subsection{Spectroscopic Data and Reduction}\label{subsec:data}

We retrieved spectroscopic data of NGC\,1275 taken with the Intermediate-dispersion Spectrograph and Imaging System (ISIS) on the 4.2--m William Herschel Telescope (WHT) from the Issac Newton Group Archive. The ISIS operates in a dual-beam mode by employing a standard D5300 dichroic mirror to form blue and red spectral channels with a crossover wavelength at 5336 \text{\AA}.  In the observation of NGC\,1275, light from the blue channel was directed to the R300B grating, which spans a usable wavelength range of 3000--5200 \text{\AA} at a spectral resolution of 7 \text{\AA} (corresponding to a velocity resolution, $\sigma_{v}$, of 171 to $297 \rm \, km \, s^{-1}$ depending on the wavelength). The light from the red channel illuminated the R316 grating, which spans a usable wavelength range of 5450-8100 \text{\AA} at a spectral resolution of 6.3 \text{\AA} (corresponding to $\sigma_{\rm{v}} = 100$--$147 \rm \, km \, s^{-1}$ depending on the wavelength). The observations used here were conducted over two nights on 2007 December 28--29  (PI: Nina Hatch, N12 + S/D) as summarized in Table\,\ref{tbl:observation}.

\begin{deluxetable}{cccccc}[htb!]
\tabletypesize{\footnotesize}
\tablewidth{0pt}
\tablecolumns{5}
\label{tbl:observation}
\tablecaption{Summary of $ISIS$ observation}
\tablehead{
\colhead{Slit} & \colhead{Date} &\colhead{grating} & \colhead{Exposure Time}}
\startdata
 1 & 2007 Dec 29 & R300B+R316R & 20 $\times$ 900\,s  \\
 2 & 2007 Dec 28 & R300B+R316R & 20 $\times$ 900\,s \\
 3 & 2007 Dec 29 & R300B+R316R & 300\,s + 3 $\times$ 600\,s \\
\enddata
\end{deluxetable}

\begin{figure*}[hbt!]
\centering
\includegraphics[width=\textwidth]{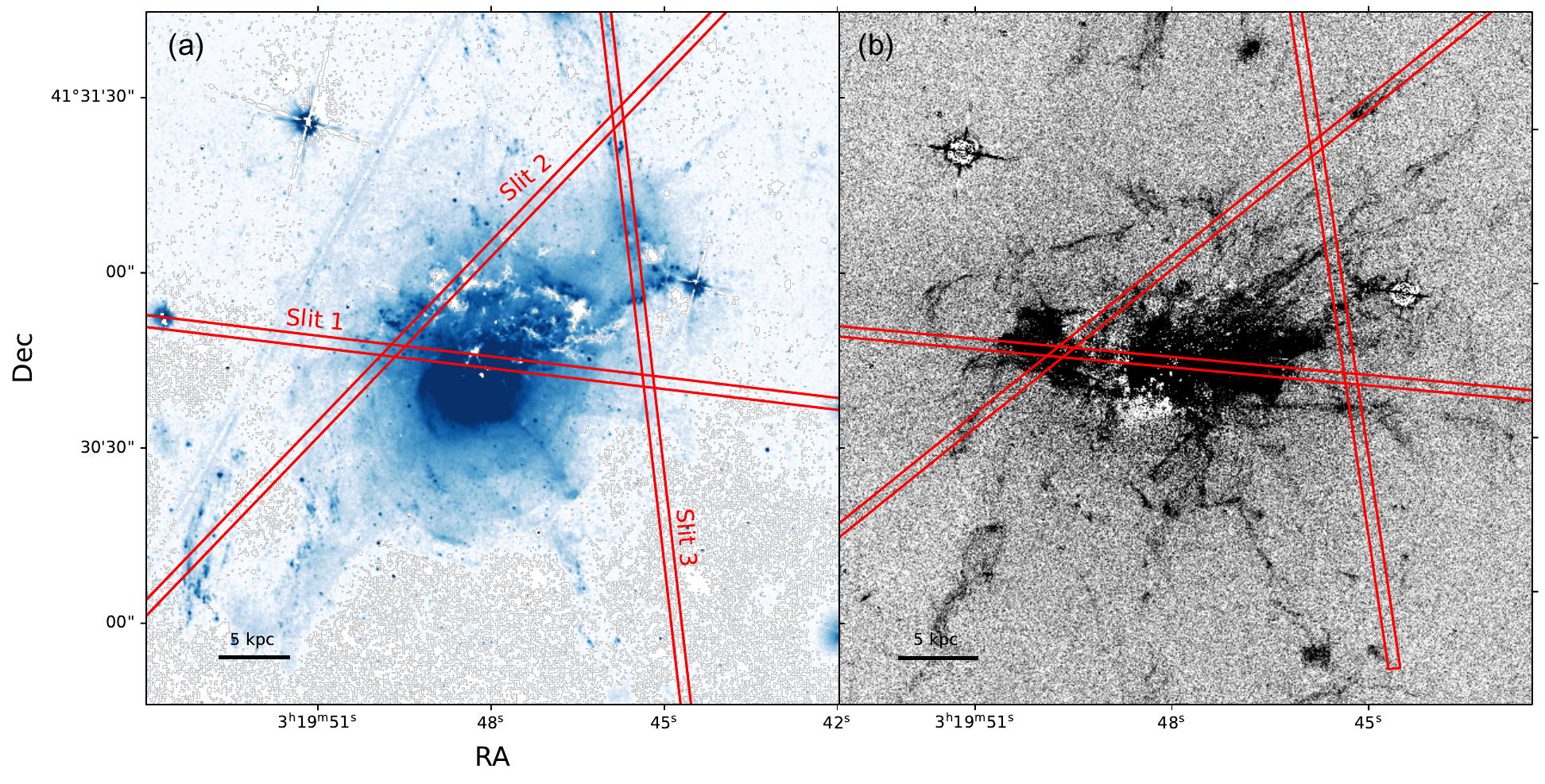}
\caption{ISIS slit positions, overlaid on: (a) a $B$-band image of NGC\,1275 after removing light from an old stellar population (see text); and (b) an HST image of the same region in H$\alpha$ + [NII].  Slit 1 cuts close to the center of NGC\,1275 at a position angle (PA) of 84$^\circ$; over the radius spanned by the extended blue light, much of this slit captures also the emission-line nebula.  Slit 2 is oriented at a PA of 132.5$^\circ$, capturing the HVS and emission-line nebula on the northern half of NGC\,1275, but is largely free of the emission-line nebula on the southern half thus capturing the extended blue light along with regions farther out.  Slit 3 does not cut through the extended blue light.}
\label{fig:slits}
\end{figure*}

Figure\;\ref{fig:slits} shows the three slit orientations along which observations of NGC\,1275 were made. 
Each slit has a width of 2".  Along both Slits 1 and 2, twenty exposures having a duration of 900\,s each were taken, yielding an integration of 5\,hr at each Slit. By contrast, along slit 3, only one 300-s exposure and three 600-s exposures were taken, yielding a shorter integration time of just 0.5\,hr. Slit 1 cuts through the inner region of NGC\,1275 at a position angle (PA) of 84$^\circ$, passing just 2\arcsec--3\arcsec\ north of the active galactic nucleus (AGN) in NGC\,1275. Over the radial range spanned by the extended blue light, much of this slit also captures either the emission-line nebula or the HVS.  Slit 2 is oriented at a PA of 132.5$^\circ$, passing closest to the center of NGC\,1275 on the north-east. Towards the north-east of center, this slit cuts through the HVS and then, further north, the emission-line nebula in NGC\,1275. South-east of center, however, the slit cuts through a region of the extended blue light that is largely free of the emission-line nebula. This portion of the slit is therefore useful for spectral analysis of the extended blue light. Slit 3 only covers regions beyond the extended blue light.

Apart from NGC\,1275, observations were made of the spectroscopic standard SP0105+625. Flats were taken at the start of observations. Finally, bias frames and CuAr+CuNe arc-lamp frames were taken on each day of the observations.

We reduced the data using the standard IRAF reduction package for long-slit spectroscopy \citep{Tody1983}. The bias (overscan) was subtracted, and flat field was applied to both the science and standard star frames. Arc-lamp frames were calibrated, and after checking through the atlas of lines, the wavelength solution applied to both the science and standard star frames. The outermost regions of each slit covering blank sky ($\sim$100\arcsec\ away from the center of NGC\,1275) were used for background sky subtraction. The spectroscopic standard star SP0105+625 was used for flux calibration. All spectra were corrected for atmospheric extinction, and de-reddened for Galactic extinction by adopting \textit{A}$_{v}$ = 0.5 based on a map of the Milky Way by \citet{Schlegel1998}. 



\begin{figure*}[hbt!]
\centering
\includegraphics[width=17cm]{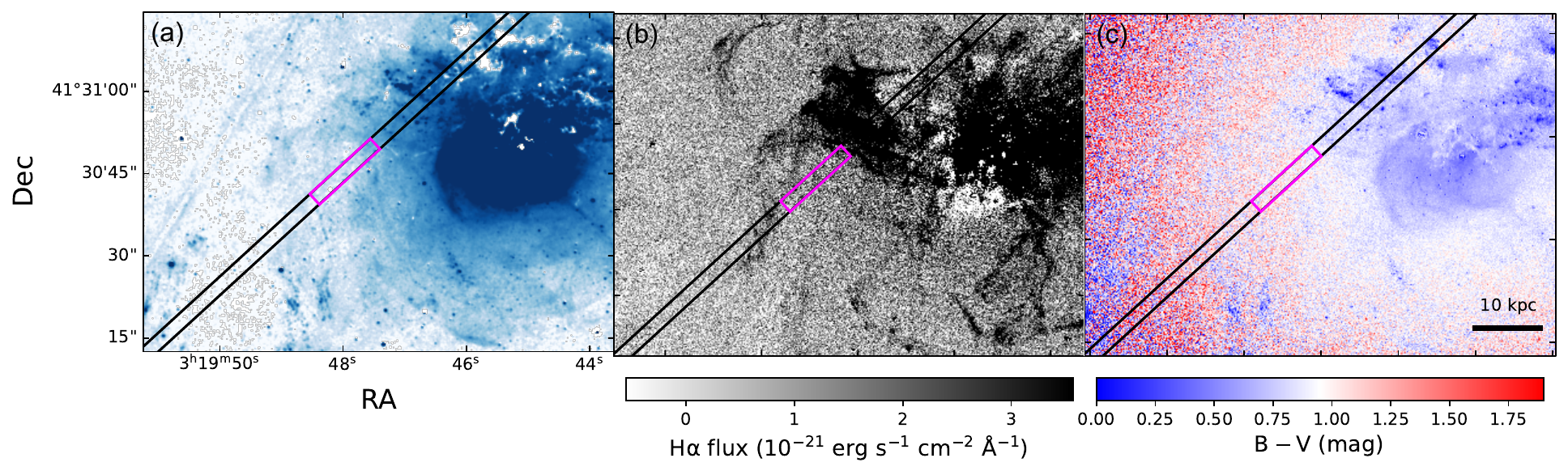}
\caption{Close-up of a portion of slit 2 that best captures the extended blue light and at the same time best avoids the emission-line nebula.  Panels (a) and (b) are the same as the corresponding panels in Fig.\,\ref{fig:slits}, and panel (c) a $B-V$ color image.  Pink box indicates the slit area over which we extracted a spectrum of the extended blue light for analysis, selected to best avoid the emission-line nebula.}
\label{fig:slit box}
\end{figure*}

\begin{figure*}[hbt!]
\centering
\includegraphics[width=17cm]{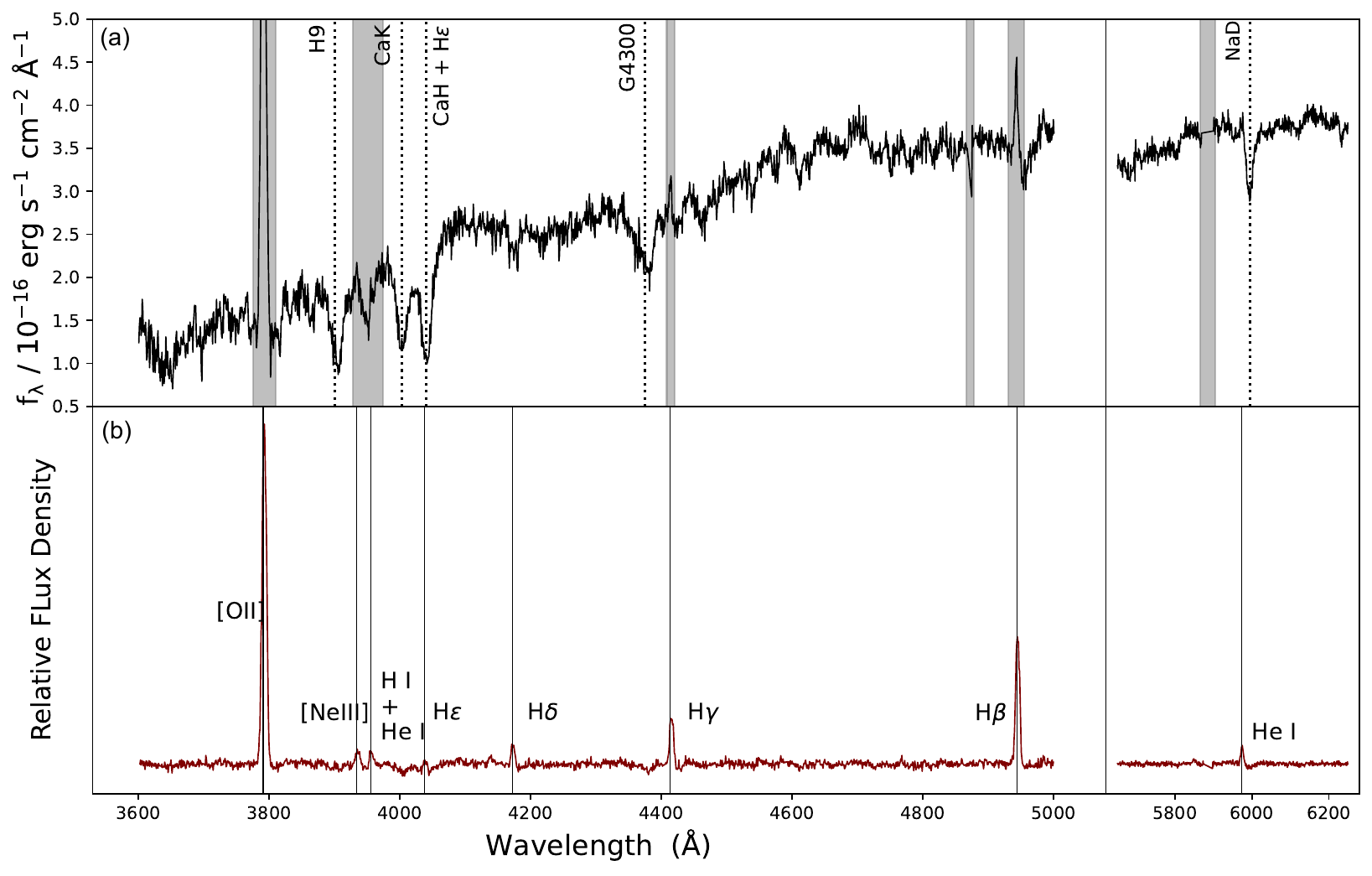}
\caption{Spectrum of: (a) extended blue light extracted over the pink box shown in Fig\;\ref{fig:slit box}; and (b) emission-line nebula, with the continuum subtracted, extracted over a slit region just north-west of the pink box.  In panel (a), grey bands indicate portions of the spectrum masked out owing to visible nebular emission lines or residuals from imperfect sky subtraction; dotted lines indicate stellar photospheric absorption lines, as labeled, for which we measured line indices. In panel (b), prominent nebular emission-lines are labeled.}
\label{fig:nebula}
\end{figure*}

Figure\;\ref{fig:slit box} focuses on where a part of Slit\,2, as delineated by the pink box, cuts through the extended blue light (Fig.\;\ref{fig:slit box}$a$,$c$) over a region that is apparently free of the emission-line nebula (Fig.\;\ref{fig:slit box}$b$).
The spectrum extracted over the region encompassed by the pink box, is shown in Figure\;\ref{fig:nebula}$a$, to be compared with a simple baseline-subtracted spectrum of the emission-line nebula shown in Figure\;\ref{fig:nebula}$b$ as extracted from an adjacent region covered by the same slit.  Although the region encompassed by the pink box was chosen to best avoid the emission-line nebula, an inspection of the spectrum taken in the blue channel nonetheless reveals detectable emission lines (most prominently [OII]$\lambda 3727 \rm \, \AA$, but also H$\beta$ and H$\gamma$). We therefore masked out regions of the spectrum displaying relatively bright nebular emission lines, namely [OII] at a rest wavelength of $3727 \rm \, \AA$, [Ne III] at $3869 \rm \, \AA$, H I + He I at $3889 \rm \, \AA$, and some other bright HI lines at $4340 \rm \, \AA$ and $4860 \rm \, \AA$, as indicated by the grey bands in Figure\;\ref{fig:nebula}$a$. Note that the selected aperture also minimizes light from the BSCs cataloged by \citet{Lim2020}. To test for any appreciable light contribution from the BSCs, we divided the aperture enclosed by the pink box in Figure\;\ref{fig:slit box} into smaller sub-apertures. We find no appreciable differences between the spectra in different sub-apertures, suggesting that any contamination from the BSCs is negligible.

\subsection{HST Broadband Data and Reduction}\label{sec:HST_images}
To provide additional diagnostics on the extended blue light, we also retrieved broadband images taken by the HST. These images were taken with the ACS/WFC camera through the $F435W$ ($B$) and $F550M$ ($V$) filters (obtained by \citealt{Fabian2008}), as well as with the WFC3/IR camera taken through the F160W ($H$) filter (as part of the Hubble Legacy Archive (DR 10.1)). Whereas the images in the $V$- and $H$-bands do not contain strong nebular emission lines, the image in the $B$ band contains the bright [OII] nebular emission line \citep{Ferland2009}. Faint traces of the nebula is indeed apparent in the $B$ band image, but their effects on broadband colors of the stellar population in NGC\,1275 have been shown to be negligible \citep{Yeung2022}. We transformed the image in the $H$-band to match the pixel scale ($0.04''$~pix$^{-1}$) of the images in $B$- and $V$-bands. We then corrected the $H$-band image for a minor misalignment in its World Coordinate System (WCS) by referencing the field Gaia stars. Finally, the images in the $B$- and $V$-bands were each convolved with separate kernels to match the angular resolution of the image in the $H$-band, for which the full-width half-maximum (FWHM) of the point-spread-function (PSF) is 0.24\arcsec\ (6 pixels). 

Accurate sky subtraction is a challenge when working with a galaxy that is very large relative to the field covered.  We noticed that the sky levels are markedly different between the northern and southern halves of both the $B$- and $V$-band images.  This difference is caused by CCD bias level jumps between the two amplifiers on the same CCD chip of the ACS/WFC camera \citep{acs_handbook}.
As our study of the continuum is restricted to the extended blue light on the southern side of NGC\,1275 (that on the northern side is complicated by the HVS), we conducted sky subtraction only on the southern half of the image.  Measurements of the sky brightness on this side is taken at a radius of $\simeq1.5'$, corresponding to $\simeq1.7$ times the effective radius of NGC\,1275 ($r_\mathrm{e}=19 $~kpc as measured by \citealt{kluge+20} in the $g'$ band).


\section{Stellar Populations} \label{sec:StellarPopulations}
We began by studying line indices (Section\,\ref{indices}) to provide insights on the nature of the stellar population responsible for the extended blue light of NGC\,1275, before proceeding to a full spectral synthesis (Section\ref{fullspecfit}) to better determine the physical properties of the stellar population involved. As we shall demonstrate, the line indices demand at least two stellar populations having very different ages, one consistent with the predominantly old stellar population characteristic of BCGs, and the other having formed relatively recently.  With this knowledge, we then fit synthetic stellar populations to the spectrum of the extended blue light using the Bayesian Analysis of Galaxies for Physical Inference and Parameter EStimation (BAGPIPES) package \citep{Carnall2018}, thus providing a more exact determination of the physical properties of the stellar populations involved. Bagpipes uses stellar population models from \citet{Chevallard2016}, for which the models themselves are based on an update to the stellar population models of \citet{Bruzual2003} (BC03).  The updated stellar population models incorporate the MILES stellar spectral library of \citet{Falcon-Barroso2011} alongside revised stellar evolutionary tracks from \citet{Bressan2012} and \citet{ Marigo2013}, applied over the age range 10$^{6}$--10$^{10.2}$\,yr and metallicity range 0.005--5 $Z_{\odot}$.  In addition, we also made use of the broadband HST images for a color analysis (Section\,\ref{subsec:broadband}), so as to provide an additional diagnostic on the stellar populations involved as well as a check on the spectral synthesis. Specifically, by subtracting away the old stellar population in NGC\,1275 using the image taken in the $H$-band, we are able to reveal in greater detail the spatial morphology of the extended blue light produced by a young stellar population as shown in Figure\,\ref{fig:slits}$a$.

\subsection{Spectral line Indices}\label{indices}

\subsubsection{Measurements}

The Lick indices provide a standardized way of measuring the strength of relatively prominent stellar absorption lines when the stellar continuum is poorly defined owing to a multitude of weak stellar absorption lines that are difficult to individually discern. A given Lick index is calculated based on three predefined wavelength intervals, corresponding individually to the absorption line of interest and its two adjacent pseudo-continuums \citep{Worthey1994}. Here, we measure line indices for the strongest absorption features observed as labeled in Figure\;\ref{fig:nebula}$a$ or as defined below (for D4000). The wavelength intervals over which these absorption features and their flanking pseudo-continuums were measured are listed in Table\,\ref{line indices}. 

As the original Lick indices do not include Balmer lines shortwards of H$\delta$, we adopt definitions for such high-order Balmer line indices -- specifically, the H9 line -- from \citet{Marcillac2006}. We do not consider Balmer lines indices longward of the H9 line owing to relatively bright nebular emission lines adjacent to or coincident with these Balmer absorption lines, as can be seen in Figure \ref{fig:nebula}$b$. Fortunately, we find no discernible nebular emission line that may contaminate the H9 stellar absorption line. Nonetheless, owing to detectable nebular emission lines near this line, we re-define the wavelength intervals of the flanking pseudo-continuums when measuring the H9 line index. We do the same also for the G4300 index (produced by CH molecule) to reduce the effects of nebular H$\gamma$ and [Ne III] emission on the flanking pseudo-continuum for this line index. For the Ca H and K indices, which are particularly sensitive to a young stellar population, we adopt the procedure described by \citet{Borghi2022}. For the D4000 discontinuity (caused by an accumulation of absorption lines by ionized metals, and measured as a ratio between average flux density in red and blue bandpasses), which also is sensitive to a young stellar population, we use the method defined by \citet{Bruzual1983}.
Finally, we use the Lick index for the NaD (Na I at 5869\,\AA) as originally defined.


\begin{deluxetable*}{ccccc}[hbt!]
\tabletypesize{\normalsize}
\tablewidth{10pt}
\label{line indices}
\tablecolumns{5}
\tablecaption{Bandpass Definitions and Measurements for Line Indices}
\tablehead{
\colhead{Index Name} & \colhead{Blue Pseudo-Continuum} & \colhead{Index} & \colhead{Red Pseudo-Continuum} & \colhead{Line Index} \\ \colhead{ } & \colhead{Bandpass (\text{\AA})} & \colhead{Bandpass (\text{\AA})} & \colhead{Bandpass (\text{\AA})} & \colhead{(\text{\AA})}}
\startdata
H9 & 3810.000--3820.000 & 3825.000--3845.000 & 3855.000--3865.000 & 6.88 $\pm$ 0.27\\
Ca H & 3950.000--3957.000 & 3962.000--3982.000 & 3987.000--3994.000 & 7.06 $\pm$ 0.23\\
Ca K & 3910.000--3917.000 & 3922.000--3940.000 & 3945.000--3952.000 & 5.12 $\pm$ 0.33\\
G4300 & 4266.375--4282.625 & 4281.375--4316.375 & 4318.875--4335.125 & 3.99 $\pm$ 0.24\\
Na5896 & 5860.625--5875.625 & 5876.875--5909.375 & 5922.125--5948.125 & 3.65 $\pm$ 0.35\\
D4000 & 3750.000--3950.000 & ... & 4050.000--4250.000 & 1.78 $\pm$ 0.01\\
\enddata

\end{deluxetable*}

We make use of the open-source software package ``indexf" \citep{Cardiel2010} for deriving all the line indices mentioned above. Indexf requires as inputs both the radial velocity of the absorption lines and its associated measurement uncertainty. We derived both these parameters using the XCSAO program, which is implemented in the RVSAO package in IRAF \citep{Kurtz1998}, by cross-correlating XCSAO template spectra for an early-type galaxy (displaying many of the same stellar photospheric absorption lines as) with the measured spectrum. In this way, from the spectrum shown in Figure\;\ref{fig:nebula}$a$, we measured a radial velocity of $5297 \pm 63 \rm \, km \, s^{-1}$. This radial velocity closely resembles that reported by \citet{Riffel2020} of 5284 $\pm$ 21 km$s^{-1}$, which is based on the CO band-head features at a wavelength of $\sim$$2.3 \, \mu$m measured within a radius of 1.\arcsec 25 from the nucleus of NGC\,1275.

\begin{figure*}[htb!]
\centering
\includegraphics[width=17cm]{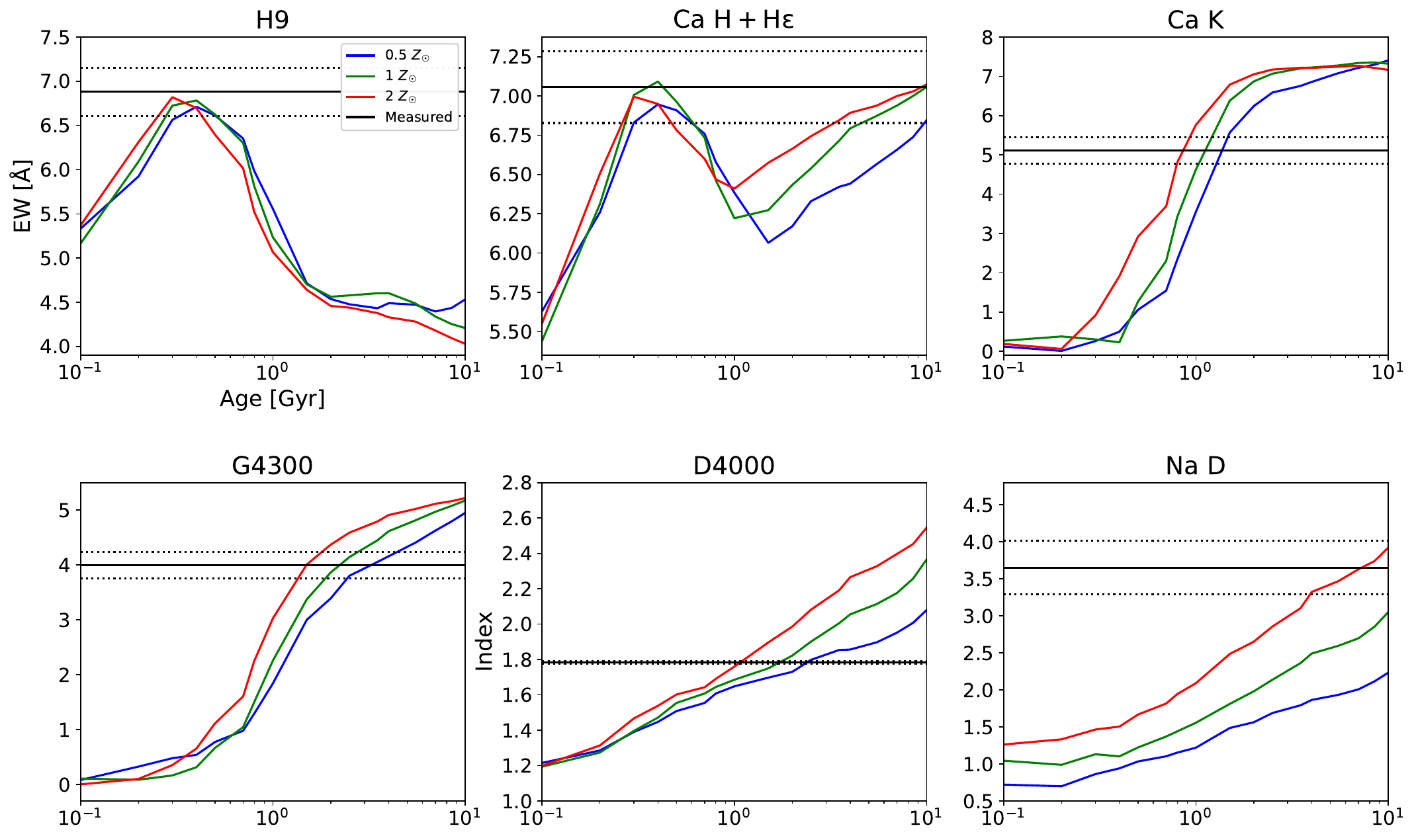}
\caption{Measured line indices indicated by black solid lines bracketed by black dashed lines at $\pm$1$\sigma$ measurement uncertainties. color curves are model-predicted dependencies of line indices with age for a coeval stellar population having a metallicity of $Z = 0.5 \rm \, Z_{\odot}$ (blue), $Z = 1.0 \rm \, Z_{\odot}$ (green), or $Z = 2 \rm \, Z_{\odot}$ (red). Reproducing H9 requires a young stellar population, whereas reproducing NaD requires an old stellar population. Remaining line indices can be reproduced by a mixture of a young and old stellar populations (see text).}
\label{fig:indices}
\end{figure*}

\subsubsection{Single SSP Model Line Indices}\label{subsec:singleindices}

To begin, we first compare the measured line indices with those predicted for just a single stellar population (SSP).  To enable such a comparison, we first generated model spectra for SSPs using BAGPIPES, after which we smoothed the model spectra to the same spectral resolution as the observed spectrum measured with ISIS. BAGPIPES makes use of the updated \citet{Bruzual2003} stellar evolutionary models, and adopts an updated Kroupa IMF \citep{Kroupa2002}. The model spectra we generated span a broad range of ages from 0.1--10 Gyr at three selected metallicities: (i) $Z = 0.5 \rm \, Z_{\odot}$, as is closely comparable to the metallicity of the ICM \citep{Grandi2001, Schmidt2002, Sanders2007} and therefore commensurate with stars produced from cooling of the ICM; (ii) $Z = \rm \, Z_{\odot}$; and (iii) $Z = 2 \rm \, Z_{\odot}$, as is characteristic of the super-solar metallicity of old stars in BCGs \citep{Loubser2009, Romeo2013}.  

We note that the (old) stellar population in massive elliptical galaxies has been found to contain elevated $\alpha$ enhancement \citep[e.g.][]{Gu2018}   We suggest that some of the discrepancies between the model and observed line indices described in Section\,\ref{subsec:cspindices} is caused by the lacks of $\alpha$ enhancement in the model spectra, as BAGPIPES uses solar-scaled models and lack the utility to implement $\alpha$ enhancement. From the model SSP spectra, we calculated model line indices in the same manner as described above for the measured line indices.  In Figure\;\ref{fig:indices}, we plot, as function of age, the values of the individual model line indices for the three selected metallicities as displayed in different colors.  The individual measured line indices are indicated by a solid black horizontal lines, bracketed by dashed black horizontal lines corresponding to the $\pm 1 \sigma$ measurement uncertainty of a given line index.

The measured H9 index is at or close to the maximal value predicted by the different model spectra, implying a young stellar population having an age of $\sim$300\,Myr. At that time, OB stars have disappeared and A stars dominate the spectrum, thus producing a peak in the H9 line index. A young stellar population with an age of $\sim$300\,Myr also is consistent with the measured Ca H + H$\epsilon$ index, although in this case degenerate with an old stellar population having an age of $\sim$10\,Gyr. On the other hand, the NaD index demands an old stellar population having an age of $\sim$10\,Gyr and super-solar metallicities \footnote{Absorption in the NaD line can also arise in the interstellar medium \citep{Conroy2014}, and can potentially therefore be enhanced by absorption through the emission-line nebula in NGC\,1275.}. Further confusing matters, the remaining line indices, Ca K, G4300, and the D4000 break, all demand a stellar population having an intermediate age of $\sim$1\,Gyr. As we will show next, the combination of a young (as demanded by the H9 line index) and old (as demanded by the NaD line index) stellar population, with no need of an additional stellar population having an intermediate age, is able to explain all the line indices shown in Figure\;\ref{fig:indices}.

\subsubsection{Composite Stellar Population Model Line Indices}\label{subsec:cspindices}

\begin{figure*}[htb!]
\centering
\includegraphics[width=\textwidth]{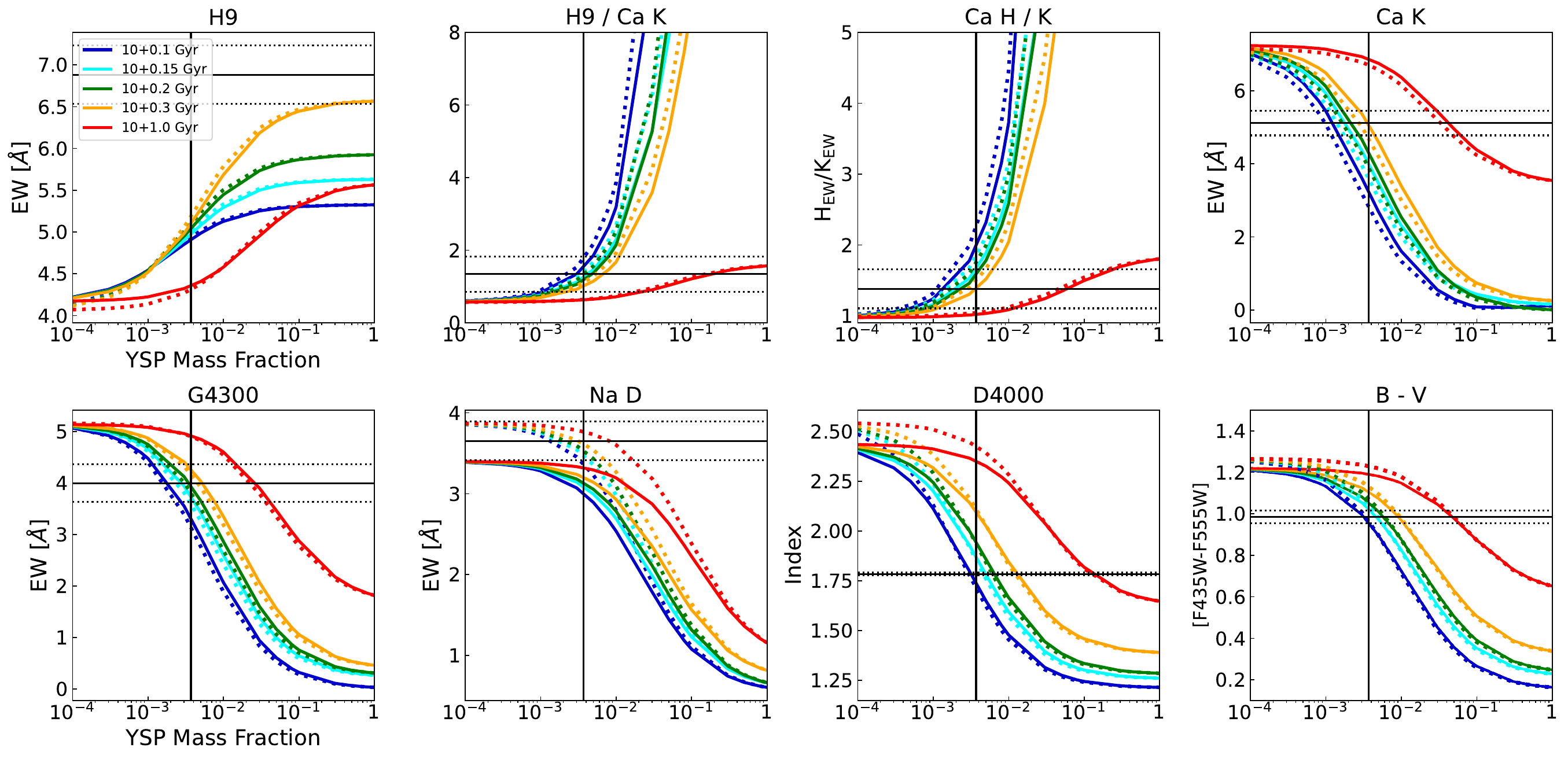}
\caption{Line indices or continuum color (lower right panel), as measured from the spectrum, indicated by black horizontal solid lines bracketed by black horizontal dashed lines at $\pm$1$\sigma$ measurement uncertainties. color curves are model predictions for a mixture of a young stellar population (YSP) having a fixed metallicity of $Z = 0.5 \rm \, Z_{\odot}$ but different ages as indicated by the legend in the upper left panel, together with an old stellar population (OSP) having an age of 10\,Gyr but different metallicities of either $Z = 1.5 \rm \, Z_{\odot}$ (dotted curves) or $Z = 2.0 \rm \, Z_{\odot}$ (solid curves).  Ordinate indicates the fractional mass of the YSP in this mixture with an OSP.  Black vertical lines indicate the YSP mass fraction from the best-fit model to the spectrum (see Fig.\,\ref{fig:doubleburstfit}), for which the YSP has an age of 0.16\,Gyr and a metallicity $Z = 0.4 \rm \, Z_{\odot}$, and the OSP an age 10\,Gyr and metallicity $Z = 1.6 \rm \, Z_{\odot}$.  For this mixture, the YSP mass fraction from the full spectral analysis is closely comparable to that required to produce the measured line indices, except for H9 (see reasons in the text), and continuum color (i.e., compare vertical black line with intersection between black horizontal lines and light-blue dotted curves).}
\label{fig:indicesCSP}
\end{figure*}

Given the inability of a coeval stellar population to reproduce all the measured line indices shown in Figure\;\ref{fig:indices}, we now consider the possibility of a composite stellar population having two very different ages and possibly also metallicities. To begin, we generate model spectra that combine two model SSPs (hereafter, 2-SSP composite model) comprising: (i) 10-Gyr stars (hereafter referred to as an old stellar population, or OSP) having super-solar metallicities (for which we consider values of 1.5 $Z_{\odot}$ and 2.0 $Z_{\odot}$), as is characteristic of the bulk stellar population in BCGs; and (ii) relatively young stars (referred to as a young stellar population, or YSP), for which we consider ages between 0.1 and 1.5 Gyr, having a sub-solar metallicity of 0.5 $Z_{\odot}$, as might be expected for stars formed from cooling of the ICM.  We note that the exact age of the OSP, or whether the OSP actually spans a range of ages, matters little so long as the stars involved have ages $\gtrsim$ 5 Gyr -- beyond which their integrated spectrum changes little with age.  The predicted line indices of the 2-SSP composite model are shown in Figure\;\ref{fig:indicesCSP} as a function of YSP mass fraction, whereby different colors correspond to different selected ages for the YSP. The solid color lines are for a metallicity 1.5 $Z_{\odot}$, and the dotted color lines for a metallicity of 2.0 $Z_{\odot}$, for the OSP, in both cases in combination with a YSP having a metallicity 0.5 $Z_{\odot}$.

Interestingly, apart from the H9 line index, all the measured line indices can be reproduced by the 2-SSP composite model given a suitable YSP mass fraction for a given YSP age.  For the two OSP metallicities considered, the YSP mass fraction required to reproduce all these line indices (apart from H9) is quite closely similar (between $\sim$$10^{-3}$--$10^{-2}$) over the YSP age range 0.1--0.3\,Gyr, increasing with YSP age to a YSP mass fraction of $\sim$$10^{-1}$ by an age of 1\,Gyr.  Note that the model predictions for the Na D line index is in better agreement with the measurements if the OSP has a metallicity of 2.0 $Z_{\odot}$ rather than 1.5 $Z_{\odot}$.

Reproducing the H9 line index with a two-SSP composite model requires both a narrower YSP age range spanning $\sim$0.3--0.5\,Gyr and a much higher YSP mass fraction than is required to reproduce the other line indices (and, even then, is in 1$\sigma$ tension with the measured H9 line index). As pointed out by \citet{Yeung2022}, the H9 index contains both the Balmer H9 and Mg\,I lines.  Mg can be enhanced by the $\alpha$ process: unfortunately, the strength of $\alpha$-enhancement cannot be altered in the SSP models\footnote{Neither can the effect of $\alpha$ enhancement be mimicked by changing the metallicity of the OSP.  As can be seen in Fig.\,5, increasing the metallicity of the OSP (at an age of $\sim$10\,Gyr) slightly decreases the strength of the H9 index.  This counterintuitive effect is caused by horizontal branch (HB) stars \citep[e.g.][]{Maraston2000}: a higher metallicity results in a weaker stellar wind (owing to a more opaque atmosphere) and hence cooler stellar atmosphere (red HB star) and therefore weaker Balmer lines, whereas a lower metallicity results in a stronger stellar wind (owing to a more transparent atmosphere) and hence warmer stellar atmosphere (blue HB star; as more of the atmosphere is stripped) and therefore stronger Balmer lines.} as implemented in Bagpipes (these models being scaled to the solar abundance as mentioned in Section\,\ref{subsec:singleindices}). Similarly, \citet{Yeung2022} find that a strong $\alpha$ enhancement is required to bring the predicted H9 index in their two-SSP composite model to (close) agreement with that observed for the the central spiral disk (see their appendix\,D). We note that the Ca\,H and K indices also are affected by $\alpha$ enhancement, as both are $\alpha$ products.  As can be seen in Figure\;\ref{fig:indicesCSP}, our two-SSP composite model also under-predicts the Ca\,K index, although not as severely as for the H9 index. To remove the effect of $\alpha$-enhancements in our analysis of line indices, we consider instead the ratio between the H9 and Ca K line indices. 
The ratio between these two line indices is shown in the relevant panel in Figure\;\ref{fig:indicesCSP}, for which we find good agreement between the predicted and observed values for our two-SSP composite model at a YSP mass fraction required to reproduce the remaining measured line indices shown in the same figure.


The ratio between the Ca II H and K lines provides another useful diagnostic for the presence of a young stellar population. 
Using the hybrid approach discussed in \citet{Borghi2022} with updated line index and pseudo-continuum definitions given in Table\,\ref{line indices}, a Ca II H to K ratio of $\gtrsim$ 1.2 indicates the presence of relatively young stars. For comparison, we measure a ratio between these two lines of 1.38 $\pm$ 0.14, providing yet another indicator for the presence of relatively young stars.

\subsection{Full Spectrum Fitting}\label{fullspecfit}

Having demonstrated that the line indices demand more than a single stellar population (Section\,\ref{subsec:singleindices}) and are seemingly consistent with two stellar populations having different ages and metallicities (Section\,\ref{subsec:cspindices}), we now consider whether the 2-SSP composite model can provide a satisfactory fit to the measured spectrum and not just the line indices. For this fitting, we again employ BAGPIPES, taking advantage of its Bayesian based approach for model parameter fitting as described by \citet{Carnall2018, Carnall2019}. In our work, we set the MULTINEST\footnote{https://github.com/JohannesBuchner/MultiNest} algorithm to 2500 live points, much larger than the default value of 400 live points so as to provide a denser sampling of the parameter space in the MCMC simulations.

\subsubsection{Simple Double Bursts of Star Formation}\label{subsec:double}

We employ the same 2-SSP composite model as described in Section\,\ref{subsec:cspindices} for fitting the measured spectrum. Like before, we do not consider the possibility of dust extinction, as we see no silhouette dust associated with the extended blue light (beyond the region covered by the HVS). In the spectral fitting, we allowed the line width to vary over a range corresponding to a velocity dispersion spanning 300--$1000 \rm \, km \, s^{-1}$.  To allow for any imperfections in the spectral calibration, we follow the standard practice of allowing a second-order Chebyshev polynomial to be multiplied with the stellar continuum. We set a narrow prior for this polynomial term so as to avoid significant changes to the spectral shape of the continuum. Table\, \ref{table:doubleburst} lists the parameters fitted along with the priors imposed on these parameters during the fit. 

\begin{deluxetable*}{llll}[]
\tabletypesize{\normalsize}
\tablewidth{40pt}
\setlength{\tabcolsep}{16pt}
\label{table:doubleburst}
\tablecolumns{4}
\tablecaption{Parameter Ranges and Priors for 2-SSP Stellar Populations}
\tablehead{
\multicolumn{1}{l}{Component} & \multicolumn{1}{l}{Parameters} & \multicolumn{1}{l}{Parameter Limits} & \multicolumn{1}{l}{Prior}}
\startdata
Global & Redshift ($z$) & 0.0176, 0.0178 & Uniform \\
\vspace{0.05cm} & Velocity Dispersion (km s$^{-1}$) & 300, 1000 & Logarithmic \\
\hline
Burst 1 & Age (Gyr) & 10 & Fixed \\
\vspace{0.05cm} & Metallicity ($\rm Z_\odot$) & 1, 2.5 & Uniform \\
\vspace{0.05cm} & Stellar Mass Formed (log\,$M_\odot$) & 0, 13 & Uniform \\
\hline
Burst 2 & Age (Gyr) & 0, 5 & Uniform \\
\vspace{0.05cm} & Metallicity ($\rm Z_\odot$) & 0.3, 2.5 & Uniform \\
\vspace{0.05cm} & Stellar Mass Formed (log\,$M_\odot$) & 0, 13 & Uniform \\
\hline
Calibration & Zeroth Order & 0.9, 1.1 & Gaussian ($\mu$ = 1, $\sigma$ = 0.05) \\
\vspace{0.05cm} & First Order & -0.1, 0.1 & Gaussian ($\mu$ = 0, $\sigma$ = 0.03)\\
\vspace{0.05cm} & Second Order & -0.1, 0.1 & Gaussian ($\mu$ = 0, $\sigma$ = 0.03)\\
\enddata
\end{deluxetable*}

\begin{deluxetable*}{llll}[]
\tabletypesize{\normalsize}
\tablewidth{40pt}
\setlength{\tabcolsep}{16pt}
\label{constantsfh}
\tablecolumns{4}
\tablecaption{Parameter Ranges and Priors for Extended+Burst Stellar Populations}
\tablehead{
\multicolumn{1}{l}{Component} & \multicolumn{1}{l}{Parameters} & \multicolumn{1}{l}{Parameter Limits} & \multicolumn{1}{l}{Prior}}
\startdata
Global & Redshift ($z$) & 0.0176, 0.0178 & Uniform \\
\vspace{0.05cm} & Velocity Dispersion (km s$^{-1}$) & 300, 1000 & Logarithmic \\
\hline
Burst & Age (Gyr) & 10 & Fixed \\
\vspace{0.05cm} & Metallicity ($\rm Z_\odot$) & 1, 2.5 & Uniform \\
\vspace{0.05cm} & Stellar Mass Formed (log\,$M_\odot$) & 0, 13 & Uniform \\
\hline
Constant SFH & Age Max (Gyr) & 0.05, 2 & Uniform \\
\vspace{0.05cm} & Age Min (Gyr) & 0.05 & Fixed \\
\vspace{0.05cm} & Metallicity ($\rm Z_\odot$) & 0.3, 2.5 & Uniform \\
\vspace{0.05cm} & Stellar Mass Formed (log\,$M_\odot$) & 0, 13  & Uniform \\
\hline
Calibration & Zeroth Order & 0.9, 1.1 & Gaussian ($\mu$ = 1, $\sigma$ = 0.05) \\
\vspace{0.05cm} & First Order & -0.1, 0.1 & Gaussian ($\mu$ = 0, $\sigma$ = 0.03)\\
\vspace{0.05cm} & Second Order & -0.1, 0.1 & Gaussian ($\mu$ = 0, $\sigma$ = 0.03)\\
\enddata

\end{deluxetable*}

\begin{figure*}[ht!]
\centering
\includegraphics[width=14.5cm]{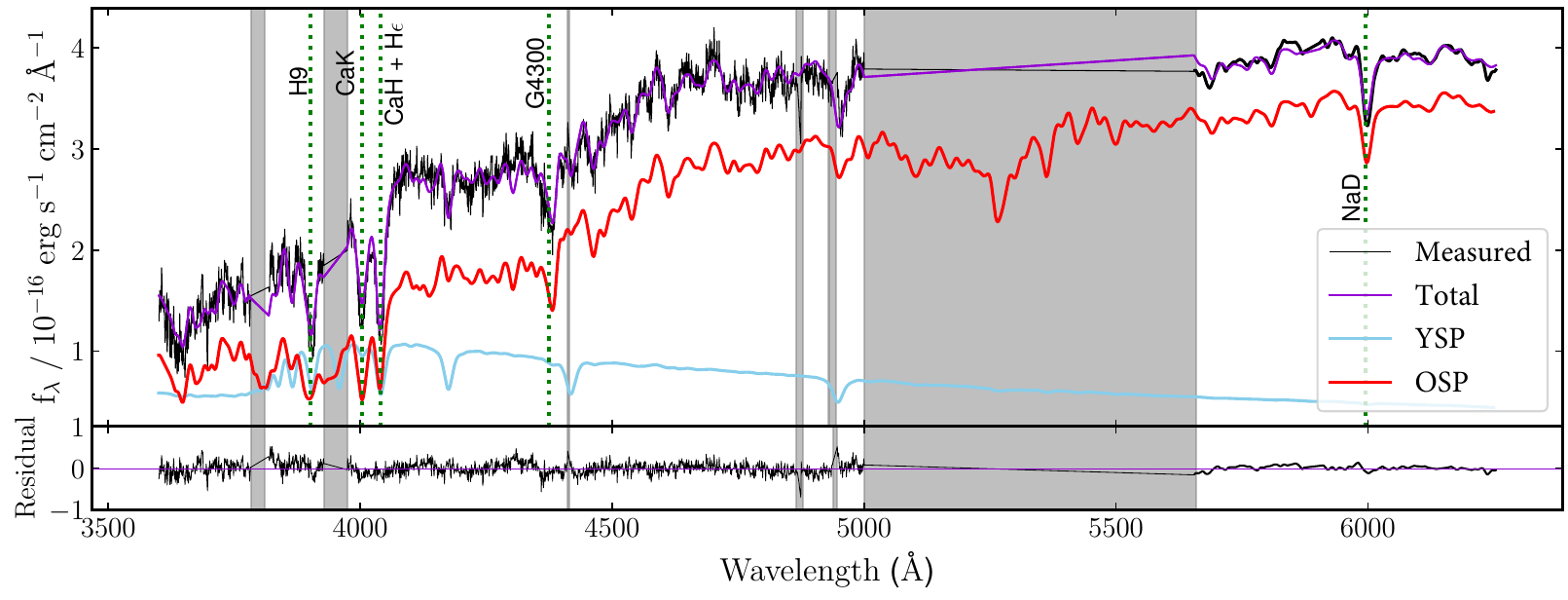}
\includegraphics[width=13cm]{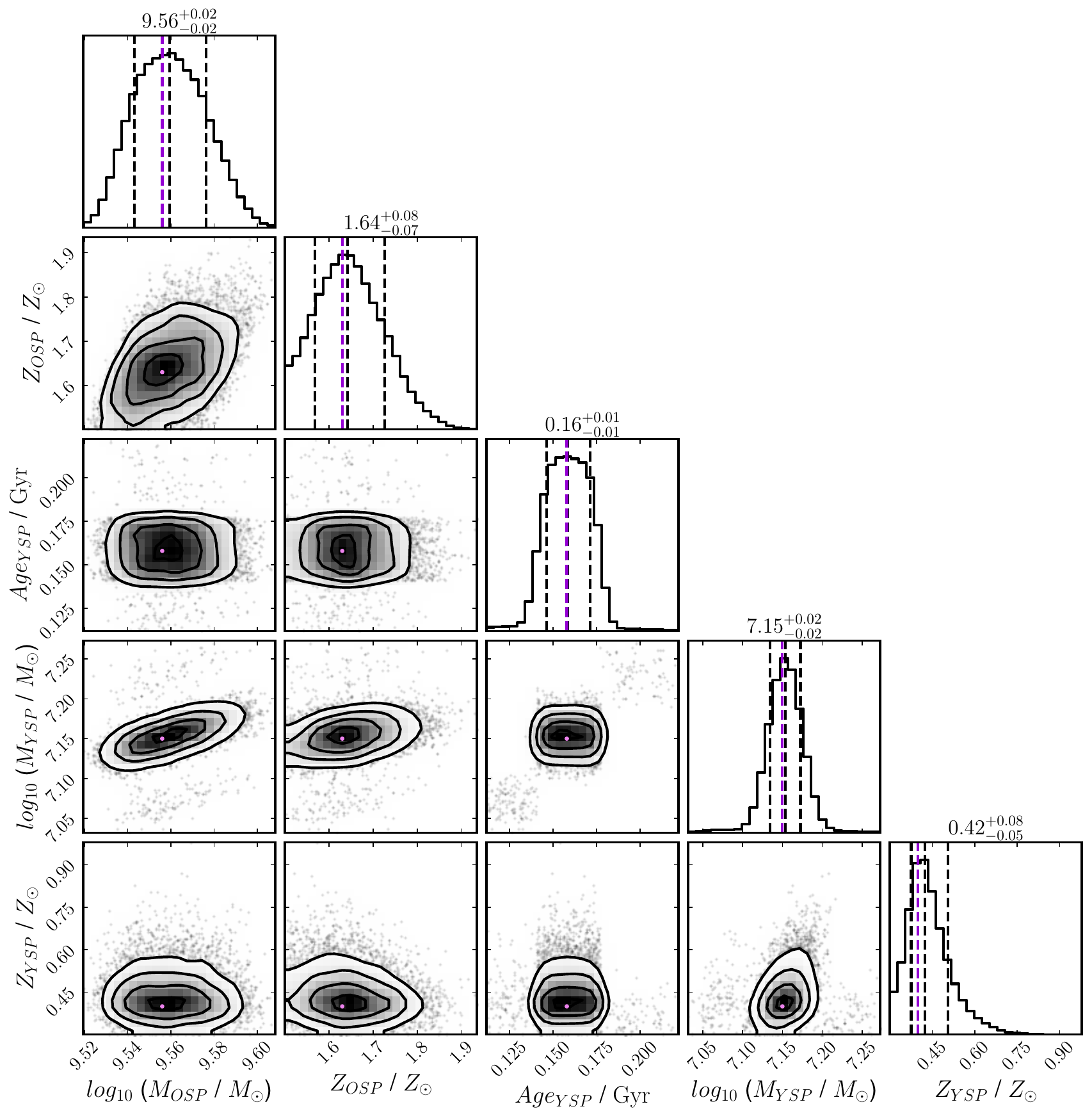}
\caption{Results of full spectral fitting for a two-SSP composite model (see text). Top panel shows measured spectrum of extended blue light (extracted over pink box in Fig.\,\ref{fig:slit box}) in black, for which grey bands indicate masked regions owing to nebular emission lines, dichroic crossover, or residuals from sky subtraction. Best-fit model spectrum (lowest $\chi^{2}$) in purple is for combination of a young stellar population (YSP), producing the spectrum in blue, and an old stellar population (OSP) having a fixed age of 10\,Gyr, producing the spectrum in red. Second panel from top shows residuals after subtracting best-fit from measured spectra. Remaining panels are 1-dimensional (1D) or 2-dimensional (2D) posterior probability distributions from the MCMC simulations.  Parameters for best-fit model are indicated by purple dashed vertical lines in the 1D plots or purple dots in the 2D plots. Black dashed vertical lines indicate, from left to right, the 16th, 50th, and 84th percentiles of the different probability distributions.}
\label{fig:doubleburstfit}
\end{figure*}

Figure\; \ref{fig:doubleburstfit} shows the best-fit 2-SSP composite model spectrum (purple) overlaid on the measured spectrum (black), for which the wavelength ranges omitted from the fit are indicated by grey bands. The omitted wavelength ranges span nebular emission lines (see Section\,\ref{subsec:data}), as well as the crossover region (the broad grey band) between the red and blue channels where the throughput is poor and artefacts prevalent (resulting in poor spectral calibration). To make clear the contribution of each stellar population to the model spectrum, we also plot in blue the contribution from the YSP and in red the contribution from the OSP. The posterior probability distributions in ages, metallicity, and stellar masses of the YSP and OSP are shown in the lower panels of Figure\; \ref{fig:doubleburstfit}. 
The black dashed vertical lines in these panels indicate (from left to right) the 16th, 50th, and 84th percentiles of the posterior probability distribution. The parameter values of the best-fit spectrum, based on least $\chi^{2}$ statistics, are indicated by the dashed purple vertical line in each panel, corresponding to a YSP having an age of $160^{+10}_{-10}$ Myr, a metallicity of $0.42^{+0.08}_{-0.05}$ $Z_{\odot}$, and a stellar mass of $10^{7.15^{+0.02}_{-0.02}}$ $M_{\odot}$ at birth. By comparison, for the OSP, which is set to have an age of 10 Gyr, the best-fit spectrum is for a metallicity of $1.64^{+0.08}_{-0.07}$ $Z_{\odot}$ and a stellar mass of $10^{9.56^{+0.02}_{-0.02}}$ $M_{\odot}$. The YSP responsible for the extended blue light therefore has a mass that is two orders of magnitude lower than that of the OSP along the same light of sight. Recall that we set priors for the metallicity of YSP and OSP as listed in Table\,\ref{table:doubleburst}. Although the OSP is required to have a metallicity of $> 1 Z_{\odot}$, the YSP is allowed to have range of metallicity spanning 0.3 to 2.5 $Z_{\odot}$. Yet, interestingly, the best-fit metallicity for the YSP is consistent with that of the ICM.

As can be seen from Figure\,\ref{fig:doubleburstfit}, the 2-SSP composite model generally reproduces most of the salient features of the measured spectrum. Consistent with the model fits to the line indices involving just a single stellar population as described in Section\,\ref{subsec:singleindices}, the Na D absorption line is largely produced by the OSP. Similarly, the G4300 and Ca K lines are produced almost entirely by the OSP, although requiring a minor contribution from the YSP. By contrast, the especially deep H9 line, as well as the Ca H + H$\epsilon$ lines, require a larger contribution from the YSP compared with the OSP. We attribute any small mismatches in fitting the H9 and Ca K lines to the effects of $\alpha$ enhancement involving the OSP as described in Section\,\ref{subsec:cspindices}.

\begin{figure*}[htp!]
\centering
\includegraphics[width=14.5cm]{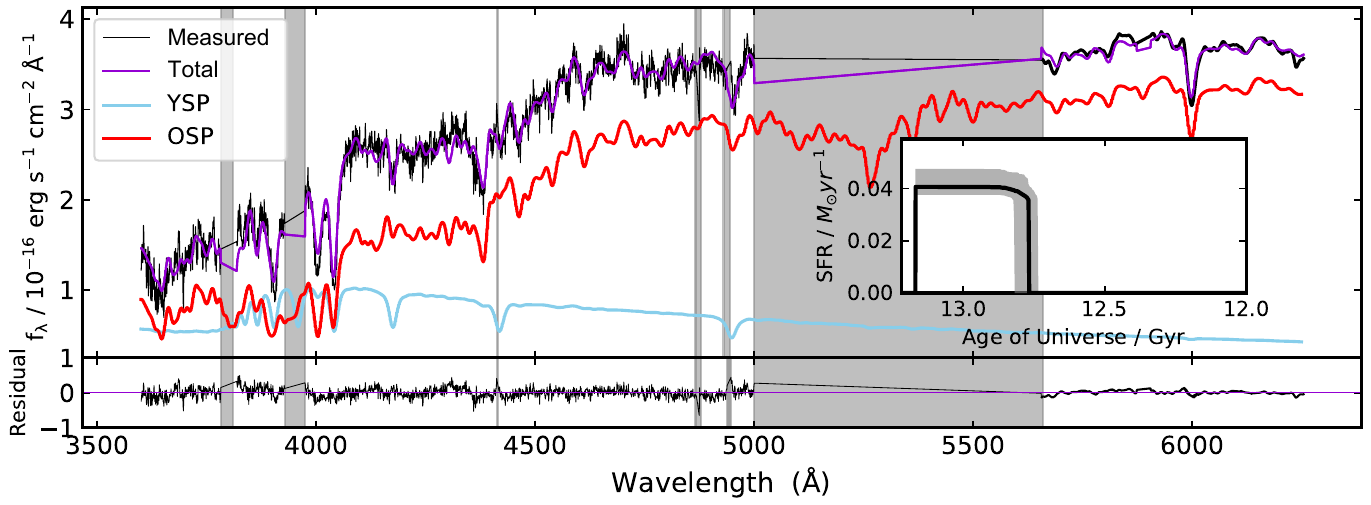}
\includegraphics[width=13cm]{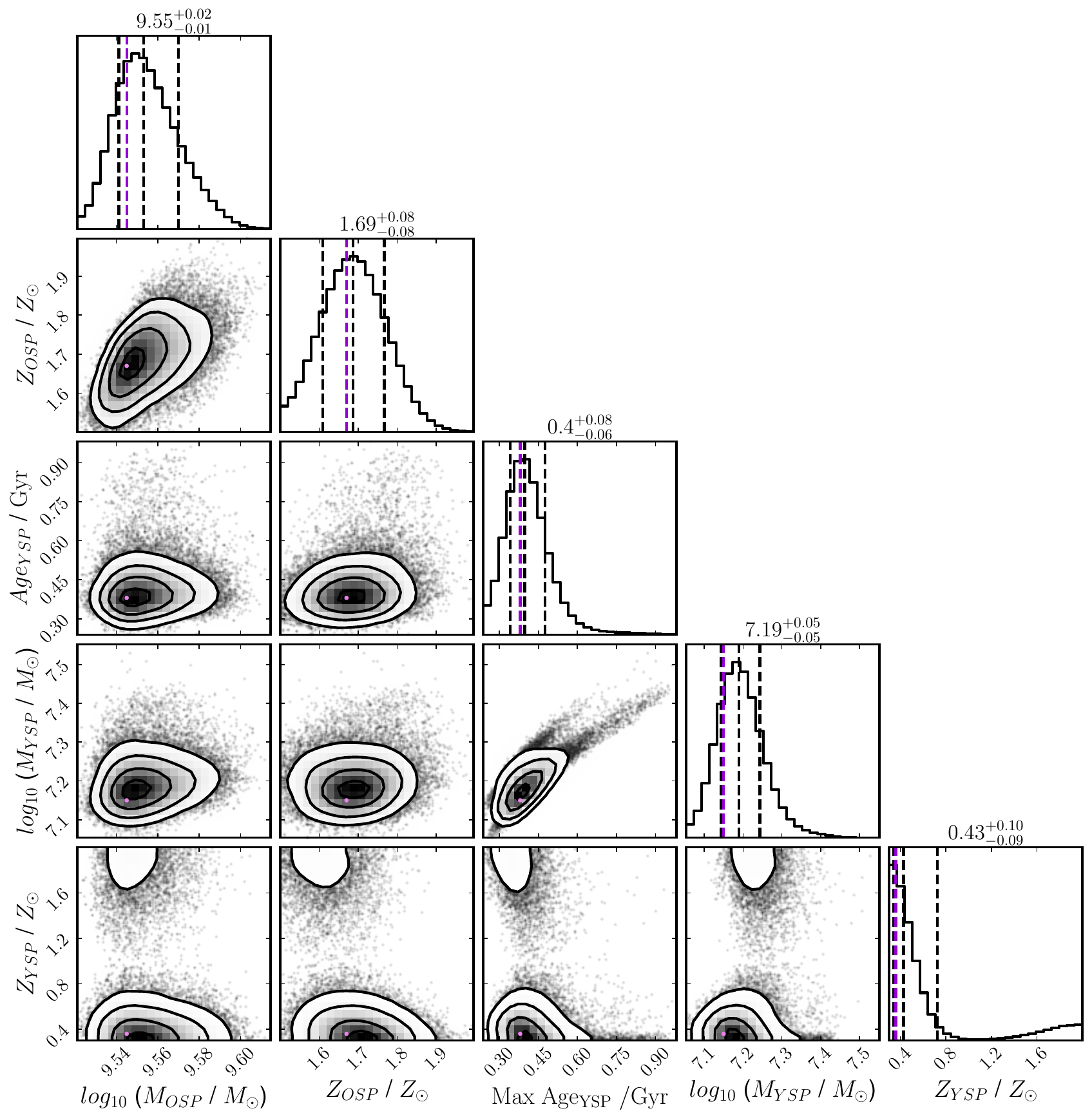}
\caption{Same as Fig\,\ref{fig:doubleburstfit}, but now for a best-fit model spectrum for a YSP formed at a constant rate over a finite duration as shown in the inset in the top panel, together with an OSP having a fixed age of 10\,Gyr. This model spectrum provides a comparably good fit as the model spectrum shown in Fig\,\ref{fig:doubleburstfit}, making it possible that the YSP spans a narrow range of ages rather than just a single age}
\label{fig:constantfit}
\end{figure*}

\subsubsection{Dependence on Adopted Star-Formation History}\label{subsec:constant}

To test whether the measured spectrum necessarily demands the YSP to have formed in a single short-duration burst, we tried two other formation histories for the YSP: (i) a constant rate of star formation over a finite duration; and (ii) a sudden burst followed by an exponential decay.  For the OSP, we adopted like before a single burst of formation 10-Gyr ago and a metallicity constrained over the range 1--$2.5 \rm \, Z_{\odot}$.   We note that a constant YSP formation rate parallels the formation history of BSCs as found by \citet{Lim2020}.  From fitting a composite model where the YSP formation is constant over time, we found that the formation of the YSP could not have continued to the present time as such a model spectrum predicts emission lines not seen in the measured spectrum, as well as a bluer continuum than is observed.  We therefore chose to terminate the YSP formation 50-Myr ago, corresponding to the timescale for the tidal disruption of star clusters at the inner regions of NGC\,1275 (see discussion in \citet{Lim2020}) -- thus providing a test of whether the extended blue light originates from the tidal disruption of star clusters formed in the same region as this light. The maximum formation age of the YSP was set to 2\,Gyr.

For the composite model adopting a constant YSP formation rate over a finite duration terminating 50-Myr ago, Figure\; \ref{fig:constantfit} (top panel) shows the best-fit model spectrum (purple) overlaid on the observed spectrum (black).  Like before, the contributions of the YSP (blue) and OSP (red) to the model spectrum are shown separately in the same panel. The maximum age of the YSP thus determined is $0.4^{+0.08}_{-0.06}$ Gyr, implying a total star formation duration of $\sim$350\,Myr (as star formation is terminated 50\,Myr ago for the reason mentioned above).  The posteriors of the model parameters are shown in the remaining panels of Figure\; \ref{fig:constantfit} below the spectrum. The metallicity of the YSP thus inferred is $0.43^{+0.10}_{-0.09}$ $Z_{\odot}$, and its total mass $7.19^{+0.06}_{-0.05}$ $M_{\odot}$. These values are closely similar to the corresponding values derived in the situation where the YSP is assumed to form in a single burst. For the OSP, the inferred metallicity is $1.69^{+0.08}_{-0.08}$ $Z_{\odot}$ and its total mass of $9.55^{+0.02}_{-0.01}$ $M_{\odot}$ at birth, again similar to the corresponding values inferred from the fit in which the YSP is assumed to have formed in a single burst. We found closely similar results for the composite model where the YSP forms in a sudden burst followed by an exponential decay, in that the decay time-scale of the best-fit model is so long as to mimic a constant YSP formation rate. Inspecting the reduced $\chi_{\nu}^{2}$ and Bayesian Information Criterion (BIC) for the different fits, we find similar goodness-of-fit for the different models of star-formation history thus allowing the possibility that the YSP formed over durations spanning hundreds of Myr, terminating several tens of Myr ago, rather than in a single short duration burst.

\subsection{Broad-Band colors}\label{subsec:broadband}

We now make use of HST images in $B$, $V$, and $H$ (as described in Section\,\ref{sec:HST_images}) to check for consistency with the spectral fitting, and to further our understanding of the temporal connection, if any, between the young stars constituting the central spiral and the extended blue light.  Recall that, in the spectral fitting, we make allowances for possible imperfections in the spectral response calibration over the large wavelength range spanned by the spectrum (see Section\,\ref{subsec:double}).  It is therefore prudent to check that the stellar populations inferred from the spectral fitting are consistent with the broad-band color of the extended blue light over the same region as the extracted spectrum (Section\,\ref{sec:bvcolors}). 
After that, we construct color images to highlight spatial variations in the color of the extended blue light over a region that is relatively free of the emission-line nebula (Section\,\ref{sec:global_colors}).  Such color variations could indicate differences in the ages of young stars at different locations, or simply spatial variations in the surface density of young stars along with that of an underlying old stellar population that give rise to different light combinations and hence resultant colors.  To check what actually contributes to the color variations, we isolated the young stars by subtracting a simple model for the light from an underlying old stellar population in the HST images (Section\,\ref{sec:extendedcolors}).  In this way, we are able to determine whether the young stars exhibit spatial variations in ages or surface densities, or both.
All the colors presented in this paper are calibrated to the Vega zero-points, and corrected for Galactic dust extinction.



\begin{figure*}[ht!]
\centering
\includegraphics[width=\textwidth]{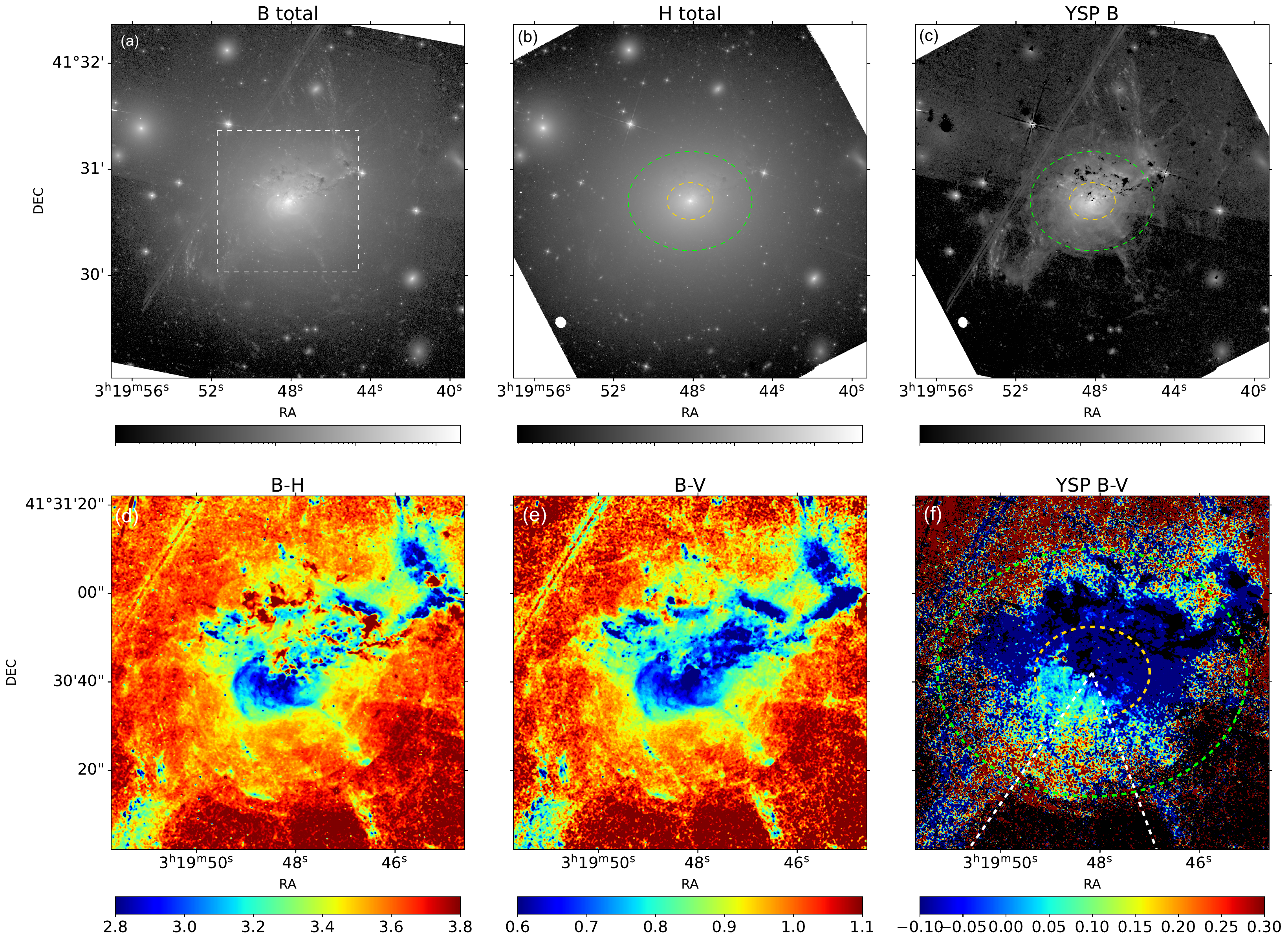}
\caption{Continuum light (log scale) and continuum colors (linear scale) of NGC\,1275 from HST images.  (a) $B$-band (F435W) image, in which central spiral disk and emission-line nebula are clearly visible.  Square denotes region encompassed by panels in lower row.  (b) $H$-band (F160W) image, tracing primarily an old stellar population (OSP). (c) $B$-band image after subtracting light from OSP (see text) thus leaving only light from a young stellar population (YSP), along with the emission-line nebula.  A multitude of arc-like features constitute, in part, the extended blue light, more clearly seen in the same images shown in Fig.\,\ref{fig:slits}$a$ and  Fig.\,\ref{fig:slit box}$a$. Ellipses in (b), (c), and (f) mark radial boundaries where there are noticeable changes in the radial color profile (refer to Fig.\,\ref{fig:colorimages2}$a$). Yellow ellipse has a radius of 13\arcsec, closely delineating boundary of central spiral disk, and green ellipse a radius of 35\arcsec, closely delineating outer boundary of extended blue light.  (d) $B-H$ and (e) $B-V$ color images, both revealing a multitude of arc-like features in the extended blue light. (f) $B-V$ color image after subtracting light from OSP (see text), highlighting the uniform continuum color of the extended blue light.  Two while lines extending from the nucleus to the south at $PA=145^{\circ}$ and $PA=240^{\circ}$ define a wedge in which we measured the radial distribution in colors, flux, and flux ratios of Fig\,\ref{fig:colorimages2}.}
\label{fig:colorimages1}
\end{figure*}

\subsubsection{$B-V$ over Slit Aperture}\label{sec:bvcolors}

Over the slit aperture (pink box in Fig.\,\ref{fig:slit box}) used to extract a spectrum for the spectral fitting, we measure $B-V  = 0.98 \pm 0.03$.  By comparison, massive elliptical galaxies that show no evidence for a young stellar population have $B-V \sim 1.2$, and hence the extended blue light is clearly bluer than that of typical massive elliptical galaxies.  Indeed, as shown below (Section\,\ref{sec:global_colors}), the main body of NGC\,1275 beyond the extended blue light has $B-V \sim 1.1$.  For the central spiral, \citet{Yeung2022} report an outwardly decreasing $B-V = 0.57$--0.86, bluer than the extended blue light over the slit aperture.  

Note that the aforementioned colors are for the combined light from a YSP and an underlying OSP, for which the resultant color depends on their relative contributions to the combined light.
With this in mind, Figure\;\ref{fig:indicesCSP} shows the predicted $B-V$ color of a 2-SSP composite model that combines a YSP and an OSP, such that the different color curves correspond to different YSP ages and a fixed OSP age of 10\,Gyr.  As would be expected, the predicted $B-V$ color becomes bluer as the YSP mass fraction increases.  The horizontal black line indicates the $B-V$ color predicted by the best-fit composite model fitted to the measured spectrum for which the YSP is assumed to have formed in a single burst, as is the case here.  Both the YSP age and its mass fraction in the best-fit composite model fitted to the measured spectrum (for which a degree of error is allowed in the continuum calibration) are closely similar to the same combination of parameters predicted to produce the measured $B-V$ color.  This check provides confidence that the stellar population fitted to the stellar photospheric absorption lines and continuum in the extracted spectrum also provides a good fit to the broad-band colors as measured by the HST over the same region.



\subsubsection{Spatial color Variations}\label{sec:global_colors}

Figure\;\ref{fig:colorimages1} shows $B$ and $H$ images of NGC\,1275, as well as a color image of this galaxy in $B-H$ along with that in $B-V$.  Two things can be immediately discerned from these color images.  First, the central spiral stands out in clear relief in both color images, appearing much bluer than the surrounding extended blue light.  Second, the region spanned by the extended blue light is not smoothly uniform in color, but exhibit arc-like features or patches that are bluer in color (appearing yellow in both the $B-H$ and $B-V$ images of Fig.\,\ref{fig:colorimages1}) than their immediate surroundings.  As mentioned above (Section\,\ref{sec:bvcolors}), the spatial variations in color need not necessarily reflect a range of ages for the young stars, but instead spatial variations in the relative contributions of a coeval YSP and an underlying OSP to the combined light and hence its resultant color.

To check for color variations over the region spanned by the extended blue light and whether the latter has a distinct outer boundary, we derived the radial color profile of NGC\,1275 in the following manner.  First, using the $H$-band image (Fig.\;\ref{fig:colorimages1}$b$), we derived the ellipticity of the galaxy and the position angle of its major axis
by fitting a double S\'ersic model to represent stars in this galaxy, along with an unresolved nuclear component to represent an AGN.  Unlike normal elliptical galaxies, two S\'ersic components were necessary to capture the global light distribution of NGC\,1275, with the inner S\'ersic component representing the region spanned by the extended blue light and the outer S\'ersic component representing the main body of the galaxy composed predominantly (if not entirely) of old stars.  The fitted parameters (ellipticity and position angle) of the outer S\'ersic component were adopted for the elliptical annuli used to generate the azimuthally-averaged colors at different distances from the center of NGC\,1275.
%

To avoid contamination by the emission-line nebula and the HVS, we chose a region that is minimally affected by the emission-line nebula as is apparent in the $B$-band image of Figure~\ref{fig:colorimages1}$a$.  As shown in Figure~\ref{fig:colorimages1}$f$, the sector chosen is a wedge with its tip centerd on the AGN in NGC\,1275 and sides at position angles of $145\degr$ and $240\degr$.  Over this sector, we masked local structures such as BSCs and globular clusters, as well as unrelated objects corresponding to field stars and galaxies (whether foreground, background, or cluster members), based on residual images constructed by subtracting the aforementioned model fits to the BCG from the original images.  As an additional step, we applied median smoothing to all images using a square box having 10 pixels on a side, and subtracted the smoothed images from the masked images so as to identify (dimmer) compact sources for further masking.

\begin{figure*}[hbt!]
\centering
\includegraphics[width=15cm]{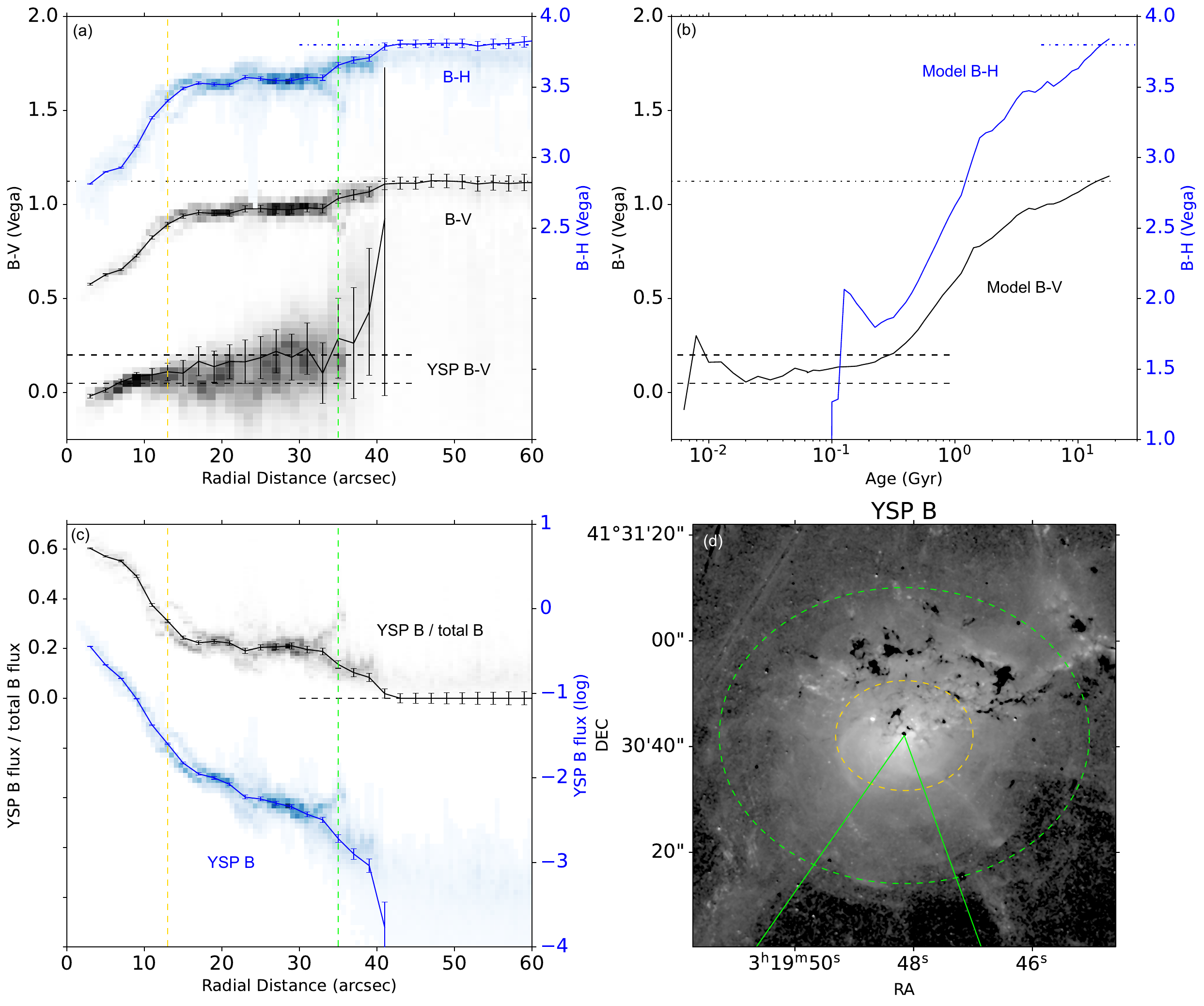}
\caption{Radial dependence in color, flux, and flux ratios along the wedge indicated in Fig.\,\ref{fig:colorimages1}$(f)$ and shown also in panel (d).  (a) Blue for $B-H$, upper black for $B-V$, and lower black for $B-V$ of young stellar population (YSP) after subtracting old stellar population (OSP).  Error bars indicate $\pm 1\sigma$ deviation from the median for all pixels within each $2''$-wide annulus along the wedge.  Shadings indicate density distributions of all individual pixels at the same radius along the wedge. Except perhaps at the innermost radii complicated by the AGN as well as the emission-line nebula in NGC\,1275, the YSP has a uniform color throughout both the central spiral disk (outer radius indicated by the yellow vertical dashed line, corresponding to yellow dashed ellipse in panel (d)) and the extended blue light (outer radius indicated by the green dashed vertical line, corresponding to green dashed ellipse in panel (d)).
(b) Model predicted color evolution of SSP in blue for $B-H$ and black for $B-V$.  In both panels (a) and (b), colors of OSP beyond the extended blue light are indicated by dash-dot horizontal lines; observed YSP B-V range is bounded by dashed horizontal lines. 
(c) Black for ratio of YSP flux in $B$-band (i.e., after subtracting OSP) to total flux in $B$-band, and blue for YSP flux in $B$-band, indicating an outward decrease in stellar surface density of the YSP.  Error bars and shadings have the same meaning as (a).  (d) B-band image after subtracting OSP, identical to that shown in Fig.\,\ref{fig:colorimages1}($f$), leaving YSP along with emission-line nebula.}
\label{fig:colorimages2}
\end{figure*}

Figure\,\ref{fig:colorimages2}$a$ shows radial profiles in $B-V$ (upper black trace) and $B-H$ (blue trace) over the wedge-shaped region shown in Figure\,\ref{fig:colorimages1}$f$ (and again in Fig.\,\ref{fig:colorimages2}$d$), computed by azimuthally averaging over elliptical annuli having an ellipticity and position angle corresponding to the main body of old stars in NGC\,1275.  The radial distance along the ordinate in both Figure~\ref{fig:colorimages2}$a$ and Figure~\ref{fig:colorimages2}$c$ corresponds to the circularized radius by deprojecting the ellipses into circles. For comparison, the slit aperture over which the spectrum shown in Figure~\ref{fig:slit box} is extracted, encompassing the region between the yellow and green ellipses in Figure\,\ref{fig:colorimages2}$d$, spans the radial range $\sim$21$\arcsec$--31$\arcsec$. Beyond distances of $r \approx 40\arcsec$ out to at least $r \approx 60\arcsec$ (the outermost distance measurable given the field of view of the HST images) from the center of NGC\,1275, the $B-V$ color does not change appreciably. Over this region, the $B-V$ color closely resembles that of (normal) massive elliptical galaxies, which as mentioned above have $B-V \sim 1.2$.  Within $r \approx 40\arcsec$, however, the $B-V$ color becomes bluer inwards until $r \approx 35\arcsec$ (green vertical dashed line in Fig.\,\ref{fig:colorimages2}$a$), and then remains constant until $r \approx 13\arcsec$ (the outermost extent of the central spiral disk, as indicated by the yellow vertical dashed line in  Fig.\,\ref{fig:colorimages2}$a$), before changing even more steeply bluer inwards.  The $B-H$ radial color profile changes in a similar manner.  The extended blue light is therefore detectable out to distance of $\sim$$40\arcsec$ (14.4\,kpc) from the center of NGC\,1275 along its major axis  -- thus spanning over two thirds of its effective radius of 19.0\,kpc. 

\subsubsection{Intrinsic colors of Young Stars}\label{sec:extendedcolors}

To determine the intrinsic color of the young stars constituting the extended blue light, and to compare this color with the intrinsic color of the young stars constituting the central spiral, we subtract the light of the underlying old stellar population that dominates the overall light from NGC\,1275 in the following manner. 
First, we select an elliptical annulus confined to the aforementioned wedge in Figure\,\ref{fig:colorimages1}$f$ to derive radial color profiles (Section\,\ref{sec:global_colors}) spanning inner and outer distances from the center of NGC\,1275 along its major axis over the range $r=44\arcsec$--$46\arcsec$. This strip is located well beyond the outermost detectable extent of the extended blue light, and therefore encloses only light from the OSP.  We scaled the brightness of NGC\,1275 in this strip as measured in $H$ to match that over the same strip as measured individually in $B$ and $V$. We then subtracted the respective scaled images in $H$ from the images in $B$ and $V$, thus leaving only light from young stars in the latter two images.  

From the OSP-subtracted $B$- and $V$-band images, we constructed the color image of Figure\,\ref{fig:colorimages1}$f$ that we refer to as the YSP $B-V$ image.  By contrast with the $B-H$ image of Figure\,\ref{fig:colorimages1}$d$ and the $B-V$ color image of Figure\,\ref{fig:colorimages1}$e$, both of which which reflect the color of the combined light from the YSP and OSP, the YSP $B-V$ image appears relatively uniform (apart from random noise fluctuations) over the region occupied by the central spiral disk and the surrounding extended blue light -- such that neither the central spiral disk nor the arc-like features in the extended blue light are visible as distinct structures.  The relatively uniform intrinsic color of young stars constituting the central spiral disk and the extended blue light is further highlighted by the lower black trace in Figure\;\ref{fig:colorimages2}$a$, showing the YSP radial $B-V$ profile azimuthally averaged over the same wedge-shaped region as the other radial color profiles shown in this panel. Within the measurement uncertainties, there is no significant change in the intrinsic color of young stars from the central spiral disk to the outermost detectable extent of extended blue light (although a weak trend of redder colors outwards cannot be ruled out).  The corresponding radial change in the azimuthally-averaged brightness -- and hence surface density -- of the apparently coeval young stars in $B$ is shown by the blue trace in Figure\;\ref{fig:colorimages2}$c$.  The brightness decreases steeply outwards over the central spiral disk, followed by a less steep outward decrease in brightness over the extended blue light up to its outermost detectable extent.

We now compare the various radial color profiles measured to the colors predicted by model SSPs having different ages as shown in Figure\;\ref{fig:colorimages2}$b$.  Within the outermost detectable extent of the extended blue light, the measured color of the YSP as indicated by the lower black trace of Figure\,\ref{fig:colorimages2}$a$ is consistent with that of a model SSP having an age of $\lesssim 300 \rm \, Myr$, as indicated by the portion of the black trace in Figure\,\ref{fig:colorimages2}$b$ bracketed by two lower black dashed horizontal lines for which there is little evolution in the predicted $B-V$ color with age. The young stars constituting the central spiral and extended blue light need not therefore be of a single age, but may span a range of ages as found for the extended blue light based on the spectral fitting (recall, from Section\,\ref{subsec:constant}, that either a single burst or a continuous formation of young stars over a limited duration, combined with an underlying OSP, provides an equally good fit to the measured spectrum in the region of the extended blue light). The total stellar mass of the YSP can be deduced from the OSP-subtracted $B$- and $V$-band images: by extrapolating over the entire annular region spanned by the extended blue light, we estimate a mass of about $1\times10^9~M_\mathrm{\sun}$ given a best-fit single YSP age of $\sim$160 Myr.

Beyond the outermost extent of the extended blue light, both the $B-V$ and $B-H$ colors shown in Figure\,\ref{fig:colorimages2}$a$ are consistent with the corresponding predicted colors of model SSPs having an age of $\sim$10\,Gyr, as indicated by the dashed-dot lines in this figure. These results justify our strategy described above for subtracting the light from the OSP so as to isolate the light from the YSP alone.

\section{Discussion}\label{sec:Discussion}



In the previous section, we showed that the spectrum of the extended blue light can be explained by stars born in a single burst $160 \pm 10 \rm \,Myr$ ago, or over a protracted duration spanning a few hundred Myr and terminating several tens of Myr ago.  If born in a single burst, their age is similar to, if not the same as, that inferred by \citet{Yeung2022} of $150 \pm 50 \rm \,Myr$ for stars in the central spiral disk -- determined by fitting a single (coeval) stellar population.
Although \citet{Yeung2022} did not try fitting stars formed over a protracted duration, it is likely that such a model would provide an equally good fit to the spectrum of the central spiral disk as we found when fitting two types of stellar population models to the spectrum of the extended blue light.  Stars that produce the extended blue light are therefore likely coeval with those in the central spiral disk.

As can be seen in Figure\,\ref{fig:slits}$a$, the extended blue light extends outwards from the outskirts of the central spiral disk.  Their close spatial as well with as temporal relationship suggest that stars in both structures were formed in the same event.  We estimate a total stellar mass (at birth) for the extended blue light of $\sim$$1 \times 10^{9} \rm \, M_{\odot}$, compared to that estimated by \citet{Yeung2022} for the central spiral disk of $\sim$$3 \times 10^{9} \rm \, M_{\odot}$.  Any explanation for the origin of the central spiral disk and extended blue light therefore needs to account for the very large mass of (molecular) gas needed to form these stars; e.g., for a Galactic-like star formation efficiency of $\sim$1\%--2\%, the total gas mass required is of order $10^{11} \rm \, M_{\odot}$.  The latter is comparable -- perhaps a factor of a few larger than -- the total mass of molecular gas inferred in NGC\,1275 at the present time based on the Galactic conversion between CO luminosity and H$_2$ gas mass \citep{Salome2011}.  What mechanism triggered an enormous mass of molecular gas in NGC\,1275 to form stars a few hundred Myr ago?  What mechanism(s) ordered the stars thus formed into a disk hosting spiral arms at the center of NGC\,1275, as well as into a multitude of arc-like features that constitute a part of the extended blue light around this disk?

\subsection{Cannibalized Gas-Rich Galaxy}


A cannibalized galaxy has been proposed by \citet{Conselice2001} and \citet{Penny2012} to give rise to the arc-like features in the extended blue light.  Such a merger also is attractive as a natural explanation for the central spiral disk, comprising the remnant of the cannibalized galaxy (perhaps itself originally a spiral galaxy).  Arc-like features (referred to sometimes as shells) are sometimes detected at the outskirts of elliptical galaxies \citep[e.g., NGC 5128, which hosts the power radio source Centaurus A;][]{Israel1998}.  Simulations suggest that about 10\% of BCGs ought to display arc-like features that correspond to the turnaround radii of relatively low-mass cluster members that have merged with the BCG \citep[e.g.,][]{kluge+20, Valenzuela2024}.  At the apogee of their orbits, stars move more slowly than at other parts of their orbits, thus bunching together to form arc-like features.

Stars that make up arc-like features in elliptical galaxies, as well as those predicted to form arc-like features in BCGs, are usually relatively old, torn from the pre-existing population of stars in the cannibalized galaxy.  By contrast, stars that make up the arc-like features in the extended blue light are relatively young, having a characteristic age (i.e., if formed in a single burst) of $\sim$160\,Myr.  As pointed out by \citet{Yeung2022}, the main difficulty in attributing the central spiral disk to a cannibalized galaxy is the large mass of relatively young stars involved, amounting to $\sim$$3 \times 10^{9} \rm \, M_{\odot}$, and hence the necessarily even larger mass of molecular gas associated with the cannibalized galaxy to form these stars -- further exacerbated by the necessity to form, in addition, stars that make up the extended blue light, having an estimated mass of $\sim$$1 \times 10^{9} \rm \, M_{\odot}$.  By the time galaxies have been slowed sufficiently by dynamical friction to merge with BCGs, much of their gas will likely have been stripped owing to ram pressure from the intracluster medium.  Even assuming that all the molecular gas accreted from a cannibalized galaxy turns into stars (i.e., at an unlikely 100\% star-formation efficiency), such a galaxy would have had to retain a mass in molecular gas of several $10^{9} \rm \, M_{\odot}$ -- equivalent to all the molecular gas in the Milky Way galaxy -- on merging with NGC\,1275 to produce the central spiral disk and extended blue light.

Alternatively, the arc-like features may have been induced by gravitational interactions with a cluster member that has penetrated deep into NGC\,1275.  Requiring a pre-existing disk of young stars around the centre of the BCG, the same interaction could have excited spiral density waves at the inner region of this disk, thus giving rise to the central spiral disk.  Such an interaction, however, is difficult to reconcile with the complete absence of any observable remnants of the since cannibalised galaxy, particularly given the short timescales involved.  Furthermore, this scenario leaves unexplained the origin of the young stars comprising (both the central spiral disk and) the extended blue light.

\subsection{Gas Cooling from Intracluster Medium}

BCGs that display luminous optical emission-line nebulae have been found to be accompanied by large masses of molecular gas as inferred from observations in CO \citep{Edge2001,Salome2003}.  Such BCGs are found to reside exclusively in clusters possessing relatively low entropies for the intracluster medium (ICM) at the cluster core \citep{Cavagnolo2008} -- see also discussion on recent revisions to our understanding of ICM entropy profiles in \citet{Levitskiy2024}.  The tight relationship between copious atomic$+$molecular gas and ICM entropy at cluster cores provides circumstantial evidence that the relatively cool gas in BCGs originates from catastrophic cooling (the portion that is not reheated by powerful jets from the active super-massive black holes in BCGs) of the ICM.  Observations in CO suggest that NGC\,1275 currently possess well over $1 \times 10^{10} \rm \, M_{\odot}$ of molecular gas \citep{Salome2006, Salome2011}, much of which is concentrated at the inner regions of the galaxy \citep{Lim2008, Salome2008}.  A similar if not somewhat larger mass of molecular gas, concentrated over a central radius of $\sim$15 kpc, would have been necessary to form stars comprising both the central spiral disk and extended blue light.


\subsubsection{Origin of Extended Blue Light}\label{subsubsec:origin}
As can be seen most clearly in both Figure\,\ref{fig:NGC1275}$f$ and Figure\,\ref{fig:slits}$a$, {some fraction of the extended blue light comprises a multitude of arc-like features.  Akin to stellar streams in the Galactic halo \citep[e.g.][]{Grillmair2016, Mateu2018, Shipp2018, Ibata2019}, we consider whether these arc-like features comprise the shredded remains of recently-formed star clusters having orbits that take them close to the center of NGC\,1275.
Thousands of relatively young star clusters have been cataloged by \citet{Lim2020} around NGC\,1275, mostly located toward the outer regions beyond the extended blue light.  These star clusters appear to have physical properties (masses and sizes) similar to globular clusters, except for their relative youth.  Referred to by \citet{Lim2020} as blue star clusters (BSCs), they have a relatively uniform distribution of ages from a few Myr to about 1\,Gyr -- unlike the extended blue light, which contains either coeval stars or have stellar ages spanning only a few hundred Myr.  The different makeup of their stellar ages makes the tidal disruptions of BSCs unlikely to be the source of the extended blue light.  Furthermore, the total mass of the detectable BSCs is only $\sim$1$\times$10$^{8}$ $ \rm \, M_{\odot}$ \citep{Lim2020}, about an order of magnitude lower than that of the stars that produce the extended blue light.

Instead, the arc-like features may be produced by an originally much larger population of star clusters of which those that remain were discovered by \citet{Holtzman1992} distributed throughout the central spiral disk.  Referred to by \citet{Lim2022} as super star clusters (SSCs) so as to distinguish them from the much more numerous but less luminous BSCs farther out, \citet{Lim2022} found that the SSCs span a relatively narrow age range of about $500 \pm \, 100 \, \rm Myr$.  They have a maximal mass of $\sim$$10^{7} \rm \, M_{\odot}$, which is about an order of magnitude larger than the maximal mass of the BSCs farther out, but comparable to the most massive globular clusters known and cataloged by \citet{Lim2020} in NGC\,1275.  Whereas the mass function of the BSCs is similar to that of the (old) globular clusters around NGC\,1275 \citep{Lim2020}, the mass function of the SSCs is much flatter \citep{Lim2022} -- suggesting that relatively low-mass SSCs have been preferentially disrupted, leaving only those sufficiently massive and compact to have (largely) escaped tidal disruption. 

If indeed the arc-like features correspond to disrupted SSCs, both should have roughly comparable brightnesses.  In Figure\,\ref{fig:arcs}, we indicate two arcs for which we used aperture photometry to estimate their brightnesses in the $B$-band.  The brighter of these two arcs, comprising the most prominent of all the arcs embedded in the diffuse blue light, has an apparent $B$-band magnitude of $m_B \simeq 21.2$.  The other, which has a brightness more characteristic of (perhaps somewhat brighter than) the other visible arcs embedded in the diffuse blue light, has $m_B \simeq 23.0$.  For comparison, as can be seen in Figure\,3 of \citet{Lim2022}, the brightest SSC has $m_B \simeq 20$ ranging to the dimmest detectable SSC of $m_B \simeq 26.5$.  Given the difference between the characteristic ages of the stars comprising the extended blue light and SSCs, the arcs should be $m_B \sim 1$ brighter than the SSCs for the same (initial) stelar mass.  Allowing for this difference, the visible arcs embedded in the diffuse blue light are therefore $m_B \sim 2.5$--4.0 magnitudes dimmer (factor of 10--40 times less massive) than the brightest (most massive) SSCs, although $m_B \sim 2.0$--3.5 magnitudes brighter (factor of 6--25 times more massive) than the dimmest (least massive) SSCs.  

\begin{figure}[htp!]
\hspace{-6mm}
\includegraphics[width=\columnwidth]{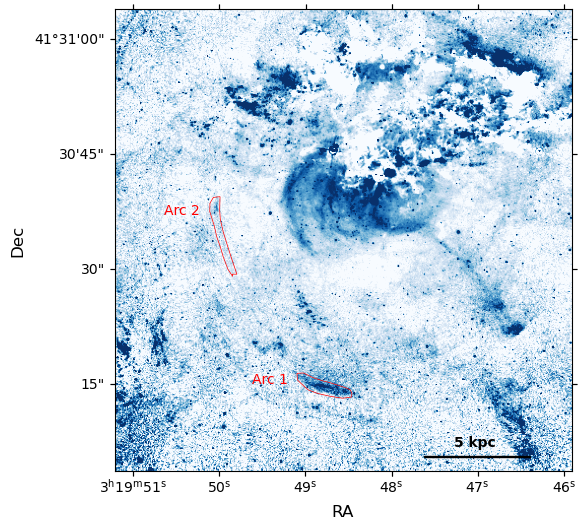}
\caption{A $B$-band image of NGC\,1275, same as in Fig\,\ref{fig:slits}a , with apertures (in red) indicating the two arcs used for the surface brightness estimation (see Discussion)}
\label{fig:arcs}
\end{figure}

Theoretical simulations by, for example, \citet{Brockamp2014} show that star clusters having the masses and sizes of globular clusters and located in the central regions of massive galaxies such as NGC\,1275 undergo a phase of rapid tidal disruption (referred to as the tidal-disruption-dominated phase) within a few hundred Myr after they are born -- comparable to the ages of stars that make up both the central spiral disk and extended blue light.  Thus, if stars that make up the extended blue light were originally born in massive star clusters, we should not be surprised to see (still coherent) stellar streams corresponding to recently disrupted star clusters embedded in this light.  We note that star clusters having lower masses (and/or larger sizes) are more easily tidally disrupted, thus flattening the mass function function of the remaining star clusters and leaving only the most massive (and/or most compact) counterparts.  As mentioned earlier, the SSCs are somewhat older than the stars in the extended blue light (and central spiral disk):\,\,we therefore suggest that the SSCs that remain formed earliest in an episode of vigorous SSC formation.  At this time, the reservoir of molecular gas available for star formation would have been at its largest and hence also the star cluster formation rate, thus producing the largest numbers of SSCs at the high-mass tail of their mass distribution.  The SSCs that formed later, having lower maximal masses, are in the process of or have recently been tidally disrupted thus leaving the arcs, or have been tidally disrupted some time ago thus leaving the smooth component of the extended blue light.

Assuming that the SSCs presently visible originally had a mass function similar to that of the BSCs, represented by a power-law mass function with slope of --2.37 $\pm$ 0.06 as measured by \citet{Lim2022}, we now estimate the total mass of this original population of SSCs.  We assume a minimal SSC mass among this original population of $\sim$$10^{4} \rm \, M_{\odot}$, comparable to the minimal mass of the BSCs farther out (imposed by the detection threshold of the images).  In this way, we estimate a total mass of $\sim$$1 \times 10^{9} \rm \, M_{\odot}$ for the original population of SSCs, similar to the estimated total stellar mass of the extended blue light.  Stars that make up the extended blue light could therefore have originated from disrupted stars clusters, of which those that remain are the especially massive SSCs discovered by \citet{Holtzman1992} distributed throughout the central spiral disk.

\subsubsection{Origin of the Central Spiral Disk}
Our estimate for the total stellar mass of the original population of SSCs, amounting to $\sim$$1 \times 10^{9} \rm \, M_{\odot}$, is highly conservative.  First, the assumed minimal mass of the SSCs is set by the detection threshold for the BSCs farther out. Second, the value estimated assumes no later episode of SSC formation following that which formed those still visible today.  In Section\,\ref{subsubsec:origin}, however, we argued that the SSCs likely formed over an extended duration, with those remaining constituting the most massive SSCs that formed earliest.  Taking both these factors into consideration, the total stellar mass of the original population of SSCs could have been considerably higher than the value estimated above of $\sim$$1 \times 10^{9} \rm \, M_{\odot}$, perhaps sufficient to give rise also to the central spiral disk (through tidal disruptions of SSCs).  The latter has an estimated stellar mass only about three times higher than the stars that produce the extended blue light.  

Alternatively, or in addition, stars comprising the central spiral disk formed from a disk of molecular gas orbiting the center of NGC\,1275.  A rotating disk of molecular gas having a radius of $\sim$100\,pc has been detected at the center of NGC\,1275 \citep{Nagai2019}, compared with a radius for the central spiral disk of $\sim$5\,kpc.

\subsubsection{AGN Stimulated Star Formation}

The formation of the central spiral disk and extended blue light required an enormous mass of molecular gas to have been concentrated in the inner region of NGC\,1275 (over a radius of $\sim$14\,kpc) at least $\sim$500\,Myr ago (the age of the SSCs).  What mechanism induced the formation of this large concentration of molecular gas, or triggered the formation of stars from this pre-existing reservoir of molecular gas?

We speculate that relativistic jets from the AGN in NGC\,1275 may have been responsible for triggering star formation from the molecular gas, if not also inducing -- to begin with -- the formation of this large concentration of molecular gas at the inner regions of NGC\,1275.  Indeed, BCGs that display luminous emission-line nebulae and vigorous star formation also preferentially display radio-luminous AGN jets \citep{Cavagnolo2008}.  NGC\,1275 displays cavities in the ICM, referred to as X-ray bubbles (largely evacuated of X-ray emitting gas), at its inner regions, for which the innermost pair of cavities spans a projected radius of $\sim$10--15\,kpc  from the center.  These X-ray bubbles are filled with relativistic electrons, detectable through their synchrotron emission at radio wavelengths.  From theoretical simulations, \citet{McNamara2016} find that the buoyant rise of such bubbles can induce thermal instabilities in the ICM to promote its catastrophic cooling, thus producing the optical emission-line nebulae in BCGs.  We speculate that compression and turbulence generated in the molecular gas by the expanding or buoyant X-ray bubbles, and/or generated as the molecular gas falls back towards the center of NGC\,1275, induce the local collapse (fragmentation) of this gas to form star clusters.  A similar process must still be occurring in the outer regions of NGC\,1275 for the molecular gas coincident with the optical emission-line nebula to form the BSCs.


The scenario we propose requires an especially vigorous episode of X-ray bubble inflation in NGC\,1275 several hundred Myr ago.  Indeed, from theoretical simulations, \citet{Falceta2010} suggest that the large X-ray bubbles currently seen at the inner regions of NGC\,1275 to have ages of 100--150\,Myr, similar to the characteristic age of stars comprising the central spiral disk and extended blue light.  Several other pairs of older X-ray bubbles have been found in NGC\,1275 detached from the center and rising buoyantly through the ICM.  Collectively, these X-ray bubbles may have induced an episode of especially vigorous episode of ICM cooling and star formation at the inner regions of NGC\,1275 that began about 500\,Myr ago, leaving the SSCs now visible throughout (or around) the central spiral disk, and for which their tidally-disrupted counterparts form the extended blue light and perhaps also contributed to stars that make up the central spiral disk.  

\section{Summary}\label{sec:Conclusion}

Unlike most elliptical galaxies in the field and probably a larger fraction of giant elliptical galaxies at the centers of galaxy clusters, NGC\,1275, the Brightest Cluster Galaxy (BCG) of the Perseus cluster, displays increasingly bluer color inwards within $\sim$14\,kpc of its center (Fig.\,\ref{fig:colorimages1}$a$).  As can be seen in both Figure\,\ref{fig:NGC1275}$f$ and Figure\,\ref{fig:slits}$a$, a bluish disk hosting spiral arms spans a radius of $\sim$5\,kpc from the center of this galaxy, surrounded by more diffuse blue light -- made up in part of a multitude of arc-like features -- extending outwards to a radius of $\sim$14\,kpc.  \citet{Yeung2022} have shown that the central spiral disk is indeed a rotating structure, comprising stars having a characteristic age of $\sim$150\,Myr and a total mass of $\sim$$3 \times 10^{9} \rm \, M_{\odot}$.  Here, we investigate the properties of the stars that produce the extended blue light, and propose a scenario for their origin that unites their properties with those of the blue and especially luminous star clusters found by \citet{Holtzman1992} projected against the central spiral disk.  The proposed scenario also suggests a revised understanding of the possible origin of the central spiral disk.


From archival long-slit spectra taken with the Intermediate dispersion Spectrograph and Imaging System (ISIS) on the 4.2-m William Herschel Telescope (WHT), we carried out a detailed analysis of the spectrum extracted from a rectangular region encompassing the extended blue light (Fig.\,\ref{fig:slit box}) that is as free as possible of the optical emission-line nebula in NGC\,1275.  First, we computed line indices for prominent stellar photospheric absorption lines apparent in the spectrum that are sensitive to different stellar ages (Fig.\,\ref{fig:indices}).  In this way, we found that the line indices demand the presence of a relatively young stellar population having ages of a few hundred Myr, along with a relatively old stellar population having ages of $\sim$10\,Gyr -- the latter as would be expected for stars that usually dominate BCGs.  
Second, from fitting model stellar populations to the extracted spectrum, we confirmed that two stellar populations are required -- and sufficient -- to reproduce the observed stellar photospheric absorption lines (Fig.\,\ref{fig:doubleburstfit}).  The older population has an age of $\sim$10\,Gyr and super-solar metallicities of $\sim$$2.0 Z_\sun$, both of which are characteristic of the old stars that dominate (in mass) BCGs.  The younger population, if coeval, has an age of $\sim$160\,Myr and sub-solar metallicities of $\sim$$0.5 Z_\sun$, similar to the metallicity of the surrounding intracluster medium (ICM).  We showed that the younger population required to reproduce the observed stellar photospheric absorption lines need not be coeval:\,\,they could have formed over a more protracted duration that, if formed at a constant rate over time, spans ages of $\sim$50--400 Myr.  Third, from an analysis of the continuum colors of the extended blue light, we found that the observed colors -- after removing the light of the old stellar population -- can be reproduced by stars that, if coeval, have an age of $\lesssim 300\, \rm Myr$ (Fig.\,\ref{fig:colorimages2}), consistent with our analyses based on both line indices and spectral syntheses.

The close temporal and spatial relationship between the extended blue light and central spiral disk suggests that stars in both structures were formed in the same event.  We estimated a total mass in stars (at birth) for the extended blue light of $\sim$$1 \times 10^{9} \rm \, M_{\odot}$, compared with that estimated by \citet{Yeung2022} for the central spiral disk of $\sim$$3 \times 10^{9} \rm \, M_{\odot}$.  Such a large stellar mass argues against the suggestion that these features trace a cannibalized galaxy \citep{Penny2012}, which would then require the latter to have retained an enormous mass of molecular gas despite being subject to intense ram pressure stripping before merging with NGC\,1275.  Instead, we suggest that the extended blue light, and possibly also the central spiral disk, are made of stars originally bound in star clusters that have been disrupted owing to strong tidal forces at the inner regions of NGC\,1275.  Such a scenario naturally explains the multitude of arc-like features that constitute a part of the extended blue light.  The relatively few star clusters that survived tidal disruption correspond to the super star clusters (SSCs, which have masses and sizes of globular clusters) discovered by \citet{Holtzman1992} projected against the central spiral disk -- as shown by \citet{Lim2022}, these SSCs have maximal masses of $\sim$$10^{7} \rm \, M_{\odot}$, similar in mass to the most massive (old) globular clusters found in NGC\,1275 \citep{Lim2020}, ages of $\sim$$500\pm100 \rm \,Myr$, and a mass function that is flatter than the even more numerous but somewhat less massive blue star clusters (BSCs; which also have masses and sizes of globular clusters) farther out.  If the initial mass function of the SSCs was similar to that of BSCs (which has a mass function similar to that of the old globular clusters in NGC\,1275), then the total mass of the original population of SSCs would have been sufficient to make up stars in the extended blue light and possibly also the central spiral disk (see the detailed arguments laid out in Section\,\ref{subsubsec:origin}).

Intensive studies of NGC\,1275 spanning nearly 70 years, beginning with \citet{Minkowski1957}, have slowly unraveled the enigmatic nature of NGC\,1275.  The High Velocity System (HVS), a dusty galaxy projected against and moving towards NGC\,1275 at a radial velocity of $\sim$$3000 \rm \, km \, s^{-1}$, has been shown to be a spiral galaxy experiencing intense ram pressure stripping on its first passage through the Perseus cluster \citep{Yu2015}.  Located no closer than 110\,kpc from NGC\,1275 \citep{Sanders2007}, the HVS is unlikely to be related to any of the peculiarities actually associated with NGC\,1275.  The spectacular optical emission-line nebula in NGC\,1275, found to be a multi-phase with an ionized counterpart in X-rays \citep{Fabian2006, Sanders2007} as well as a molecular counterpart in both H$_2$\,1-0~S(1) \citep{Lim2012} and CO \citep{Salome2006, Salome2008, Salome2011, Lim2008, Ho2009}, most likely originates from cooling of the ICM.  The energy source that powers this nebula is most likely to be the X-ray-emitting electrons constituting the ICM that penetrate into the nebular gas \citep{Ferland2009, Fabian2011}.  The nebula, primarily at its outskirts, has formed hundreds of progenitor globular clusters at a nearly constant rate over, at least, the past $\sim$1\,Gyr \citep{Lim2020}, thus helping explain the enormous population of globular clusters in this galaxy.  In this paper, we suggest that, about $500 \rm \, Myr$ ago, X-ray bubbles inflated by radio jets from the AGN in NGC\,1275 triggered instabilities in an enormous mass of molecular gas at the inner regions of this galaxy to form the SSCs discovered by \citet{Holtzman1992}, and possibly also induced vigorous cooling of the ICM from which these SSCs formed.  The vast majority of these SSCs were disrupted by strong tidal forces in the inner regions of NGC\,1275 to give rise to the extended blue light, within which a multitude of arc-like features trace numerous disrupted star clusters, and possibly also a rotating disk hosting spiral arms at the center of NGC\,1275.  Why the extended blue light extending out to $\sim$14\,kpc from the center of NGC\,1275 appears to comprise a disk structure in its entirety, hosting spiral arms within its inner $\sim$5\,kpc, is a mystery to be solved in a future study.

\begin{acknowledgments}
\indent We thank the referee for their constructive feedback that resulted in a significant improvement to this paper. We thank Michael Yeung for useful discussions on this work and Emily Wong for providing BSCs and GC catalogs. J. L. and A. L. acknowledge support from the Research Grants Council of Hong Kong through the General Research Fund 17300620.  Y.O. acknowledges the support by the Ministry of Science and Technology (MOST) of Taiwan through the grant MOST 109-2112-M-001-021-. This paper makes use of data obtained from the Isaac Newton Group of Telescopes Archive which is maintained as part of the CASU Astronomical Data center at the Institute of Astronomy, Cambridge. This research also employed observations made with the NASA/ESA Hubble Space Telescope and made use of archival data from the Hubble Legacy Archive, which is a collaboration between the Space Telescope Science Institute (STScI/NASA), the Space Telescope European Coordinating Facility (STECF/ESAC/ESA) and the Canadian Astronomy Data center (CADC/NRC/CSA).
\end{acknowledgments}

\vspace{5mm}
\facilities{HST(ACS WFC,WFC3 IR), WHT(ISIS)}


\software{astropy \citep{astropy:2013, astropy:2018, astropy:2022},  
           BAGPIPES \citep{Carnall2018},
           IMFIT \citep{imfit},
           IRAF \citep{Tody1983, Tody1993},
           INDEXF \citep{Cardiel2010},
           RVSAO \citep{Kurtz1998}
           MULTINEST \citep{Feroz2008, Feroz2009, Feroz2019}}





\bibliography{sample701}{}
\bibliographystyle{aasjournalv7}



\end{document}